\documentclass[aps, prl, showpacs,twocolumn,longbibliography]{revtex4-1}
\usepackage{bm, amsmath, amsfonts, amssymb, braket,mathtools,upgreek,amsthm}

\usepackage{multirow}
\usepackage{float}
\usepackage{color}
\usepackage{comment}
\usepackage{subfigure}
\usepackage{lipsum}
\usepackage[colorlinks,urlcolor=blue,citecolor=blue,linkcolor=black]{hyperref}

\begin{document}

\newcommand{\ii}{\text{i}}
\newcommand{\U}{U}
\newcommand{\V}{V}
\newcommand{\BZ}{\left[ 0, 2\pi \right]}
\newcommand{\CZ}{\left[ 0, 1 \right]}

\title{Dynamic skin effects in non-Hermitian systems}

\author{Haoshu Li}
    \email{lihaoshu@mail.ustc.edu.cn}
    \affiliation{Department of Modern Physics, University of Science and Technology of China, Hefei 230026, China}

\author{Shaolong Wan}
    \email{slwan@ustc.edu.cn}

    \affiliation{Department of Modern Physics, University of Science and Technology of China, Hefei 230026, China}

\begin{abstract}
    We study the time evolution processes of non-Hermitian systems under the open boundary condition and confirm that the unconventional reflection, dubbed the dynamical skin effect, exists in non-Hermitian systems analytically, and unveil the mechanism of its formation, which is caused by both the non-Hermitian skin effect and the Hermitian wave packet spreading. Furthermore, we find that in contrast to the uniform speed motion in Hermitian situations, the Gaussian wave packet can be accelerated and amplified during its time evolution in non-Hermitian systems. This additional motion is found to be responsible for the dynamic skin effect.
\end{abstract}

\maketitle

\emph{Introduction.---}In solid physics, one of the fundamental principles is the Bloch theorem, which solves the problem of finding energy levels in translation invariant systems. However, in the real world, translation symmetry is always violated at the materials' boundaries. Fortunately, the bulk energy levels of Hermitian systems solved under the open boundary condition (OBC) are nearly identical to those solved under the periodic boundary condition (PBC) \cite{10.2307/4145217}. This feature is broken down in non-Hermitian systems, where it is discovered that when the boundary condition is changed, the energy levels shift substantially, and a large number of energy eigenstates under OBC are located at the boundary. This phenomenon is defined as the non-Hermitian skin effect (NHSE) \cite{yao2018, yao20182, PhysRevLett.116.133903, PhysRevB.97.121401, PhysRevLett.121.026808, PhysRevX.8.031079, Thomale2019, londhi2019, song2019, PhysRevResearch.1.023013, PhysRevLett.124.056802, origin2020}. A complete theory called the non-Bloch band theory \cite{yao2018, yao20182, yokomizo2019, PhysRevLett.125.226402, PhysRevB.100.035102, PhysRevLett.124.066602, PhysRevB.101.195147, PhysRevLett.123.246801, PhysRevB.102.085151, PhysRevLett.125.186802, RevModPhys.93.015005, doi:10.1080/00018732.2021.1876991} has been developed to describe NHSE and the energy levels under OBC. NHSE is linked to chiral damping in open quantum systems \cite{song2019}, signal amplification \cite{Wanjura2020, PhysRevX.5.021025, Ranzani_2015, PhysRevLett.122.143901, PhysRevX.3.031001, PhysRevLett.112.167701, PhysRevX.5.041020, Jalas2013, Feng729, PhysRevApplied.10.047001, PhysRevX.7.031001, PhysRevB.103.L241408, PhysRevB.105.045122}, and spatially growing Green's functions \cite{ PhysRevLett.126.216407, PhysRevB.103.195157, PhysRevB.105.045122}, as well as quantized responses \cite{Li2021-1, Li2021-2}.

Recent research has revealed that non-Hermitian dynamics also exhibit intriguing properties in one-dimensional systems \cite{song2019, PhysRevLett.128.120401, PhysRevB.104.125435, sticky}, which are connected to the NHSE. From the viewpoint of non-Hermitian dynamics, the reflection process at the boundary is related to NHSE since NHSE cannot exist without the boundary. It is found numerically that in the reflection process of non-Hermitian systems with the NHSE, the reflected velocity of the wave packet is different from the conventional value. We dub this phenomenon the dynamic skin effect. Although the numerical evidence is clear, a theory about the dynamic skin effect or the reflection process in non-Hermitian systems is still absent.

In this Letter, we present an unambiguous mechanism for the formation of the dynamic skin effect, and we discover that it is related to both the NHSE and wave packet spreading. In addition, we find a different feature in non-Hermitian dynamics: Wave packets in non-Hermitian systems can be accelerated or decelerated in comparison to their Hermitian counterparts. It is an interesting discovery from both numerical and theoretical perspectives.

We begin with the one-dimensional Hatano-Nelson (HN) model \cite{hn}. It is discovered that, in addition to the initial velocity of the wave packet, the NHSE and wave packet spreading cause additional motion. In the HN model, this additional motion is the uniform acceleration or deceleration.

\emph{Model.---}The continuous version of the HN model has the Hamiltonian
\begin{align}
    H = -\frac{\hbar^2 \nabla^2}{2m} + b \nabla + E_0,
    \label{eq: conHNMain}
\end{align}
where $\nabla = \frac{d}{dx}$ in one-dimensional systems, the constant energy level $E_0 = - \frac{b^2 m}{2}$ for our convenience and we set $\hbar=1$ and $b \in \mathbb{R}$. The system is placed in a one-dimensional infinite potential well of length $L$. As a result, the coordinate $x$ satisfies $x \in [0, L] = \Omega$ and the boundary condition must be the Dirichlet boundary condition 
\begin{align}
    \psi |_{\partial \Omega} = 0.
\end{align}
The Hamiltonian $H$ is Hermitian when $b = 0$, else it is non-Hermitian. Under PBC, the energy level can be calculated in $k$ space as $E(k) = \frac{k^2}{2m} + i b k + E_0$.

The equation of motion is the effective Schr{\"o}dinger equation 
\begin{align}
    i \partial_t \psi = H \psi,
\end{align}
which can be realized by the non-unitary quantum walks \cite{PhysRevLett.119.130501, PhysRevA.98.063847, PhysRevLett.122.020501, PhysRevLett.123.230401, Xiao2020}, ultracold atom systems controlled by light \cite{Ren2022ChiralCO, sticky} and coupled waveguides \cite{Rechtsman2013, Weimann2017}.
The Gaussian wave packet is considered as the initial state
\begin{align}
    \braket{x | \psi(0)} = \frac{1}{(2 \pi \sigma^2)^{\frac{1}{4}}} e^{-\frac{x^2}{4 \sigma^2} + i k_0 x}.
\end{align}
It has a nonzero initial speed $v_0 = \textrm{Re}\left[\frac{\partial E(k)}{\partial k} \vert_{k_0}\right] = \frac{k_0}{m}$.

\begin{figure} 
    \centering 
    \includegraphics[width=0.45\columnwidth]{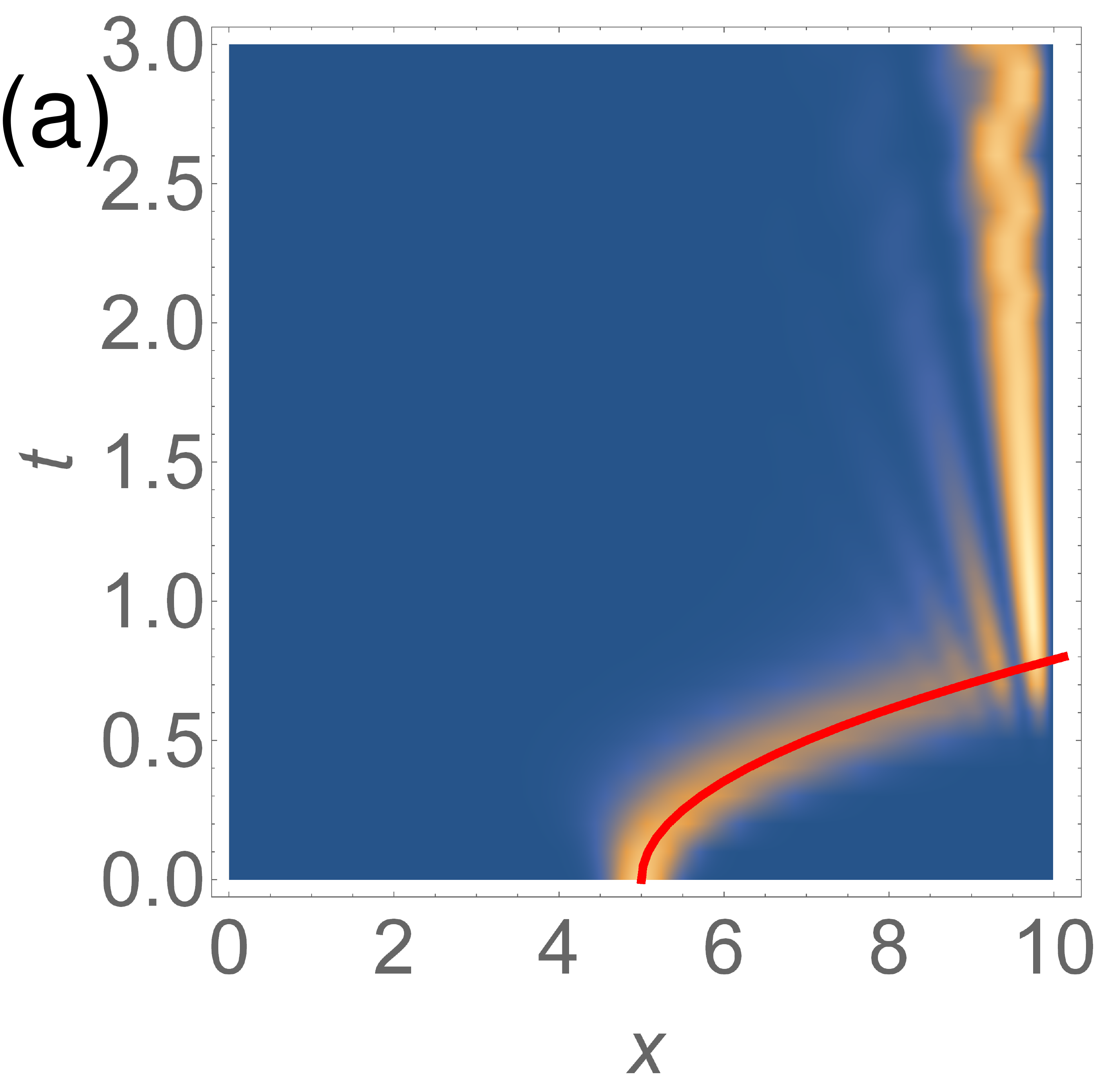}
    \includegraphics[width=0.45\columnwidth]{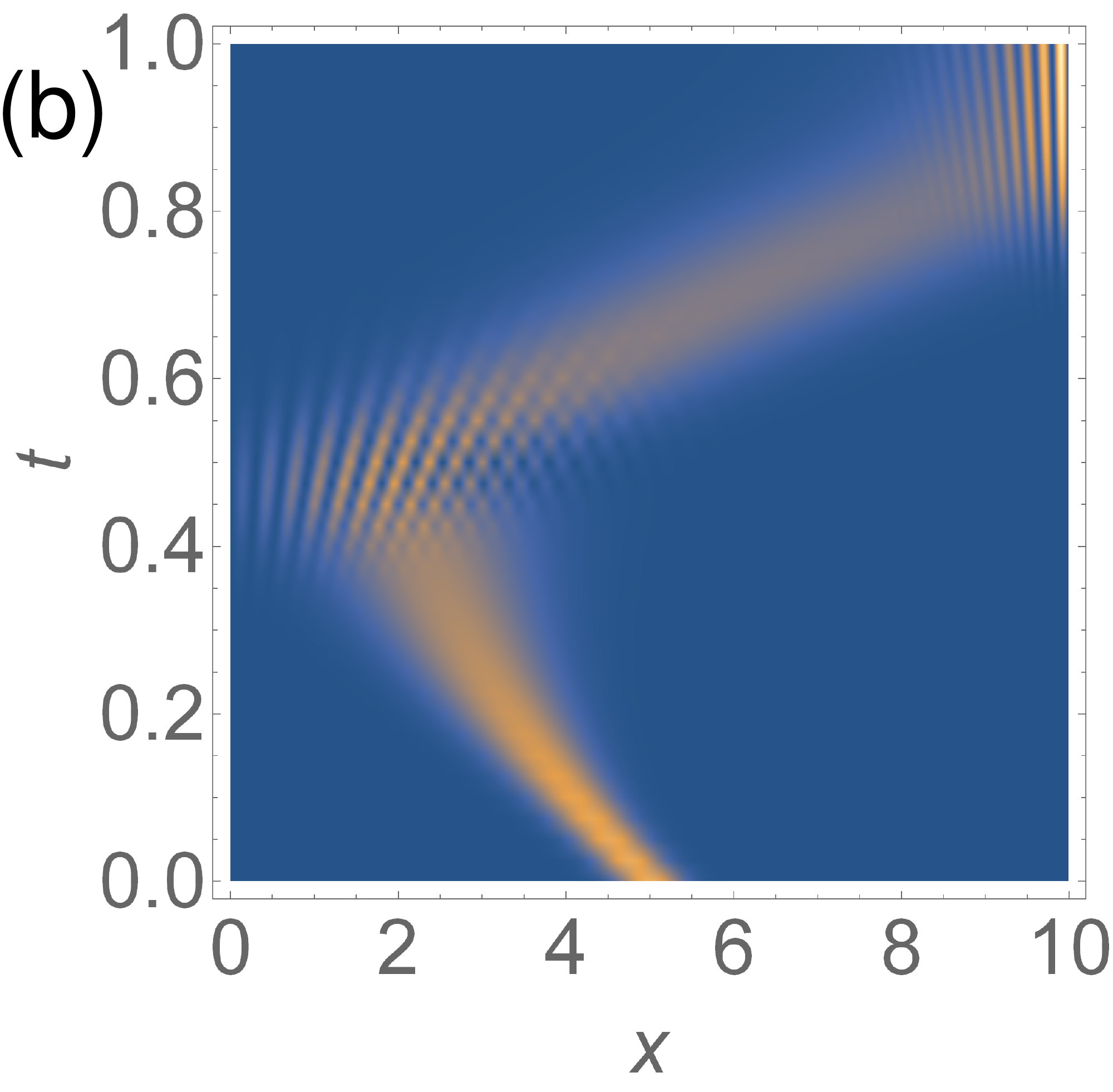}
    \includegraphics[width=0.45\columnwidth]{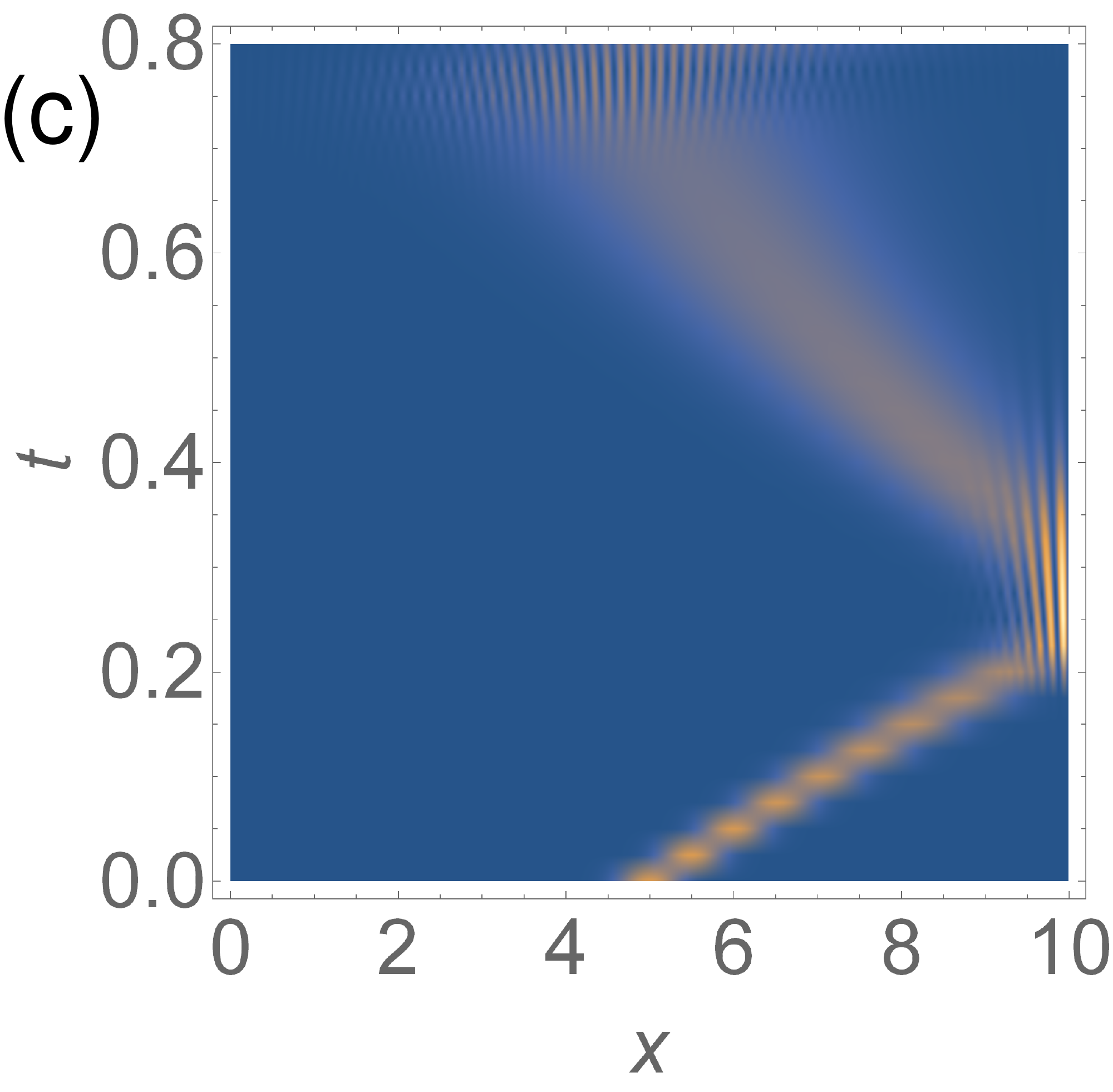}
    \includegraphics[width=0.45\columnwidth]{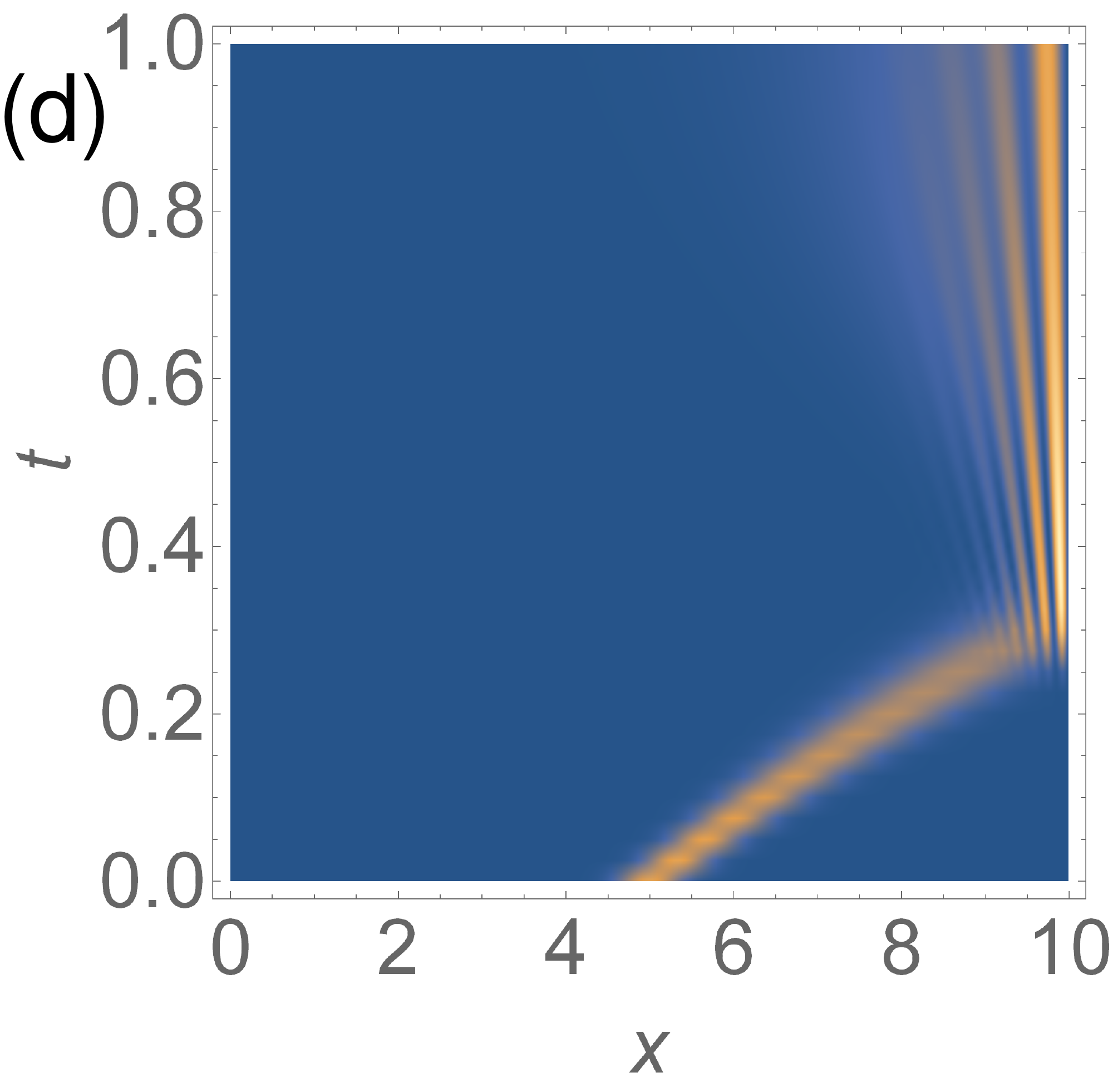}
    \caption[]{Numerical simulation of the probability density $|\braket{x | \psi(t)}|^2$, it is plotted as a function of $x$ and $t$ \cite{normalization_note}. We use the values $m = b = 1$, $\sigma=0.25$ and $L = 10$. (a) $k_0=0$: The wave packet is uniformly accelerated before it reaches the boundary. The analytical result of the wave packet trajectory is shown as the red curve.  (b) $k_0=-10$: The wave packet moves from right to left in the beginning. The velocity of the incident wave packet $|v_{\textrm{in}}|$ is smaller than the velocity of the reflected wave packet $|v_{\textrm{ref}}|$. (c) $k_0=20$: The wave packet moves from left to right in the begining. The velocity of the incident wave packet $|v_{\textrm{in}}|$ is larger than the velocity of the reflected wave packet $|v_{\textrm{ref}}|$. (d) $k_0 = 13$: The initial velocity of the wave packet is close to the critical velocity between the reflection and the non-reflection cases. In this case, the wave packet is slightly reflected. }  
    \label{fig: x-t_density}
\end{figure}

Using the finite difference method, the time evolution of the initial state $\ket{\psi(0)}$ can be numerically simulated, as shown in Fig.~\ref{fig: x-t_density}. In the simulation, we use $m = b = 1$ and $L = 10$, and more information about our numerical method can be found in the Supplemental Material (SM) \cite{supplemental}.

First, let us consider the $k_0=0$ case, in which the initial velocity of the wave packet $v_0 = 0$. According to our knowledge of the Hermitian dynamics, if the Hamiltonian is Hermitian, the wave packet will stay at its original location. However, as seen in Fig.~\ref{fig: x-t_density}(a), the wave packet is uniformly accelerated before it reaches the boundary. We ask the following question: Why does this additional motion of the wave packet happen? The answer will reveal that the NHSE is not the only factor contributing to this motion. It is also worth noting that once the wave packet hits the boundary, it is not reflected but instead remains there.

\begin{figure} 
    \centering
    \includegraphics[width=0.45\columnwidth]{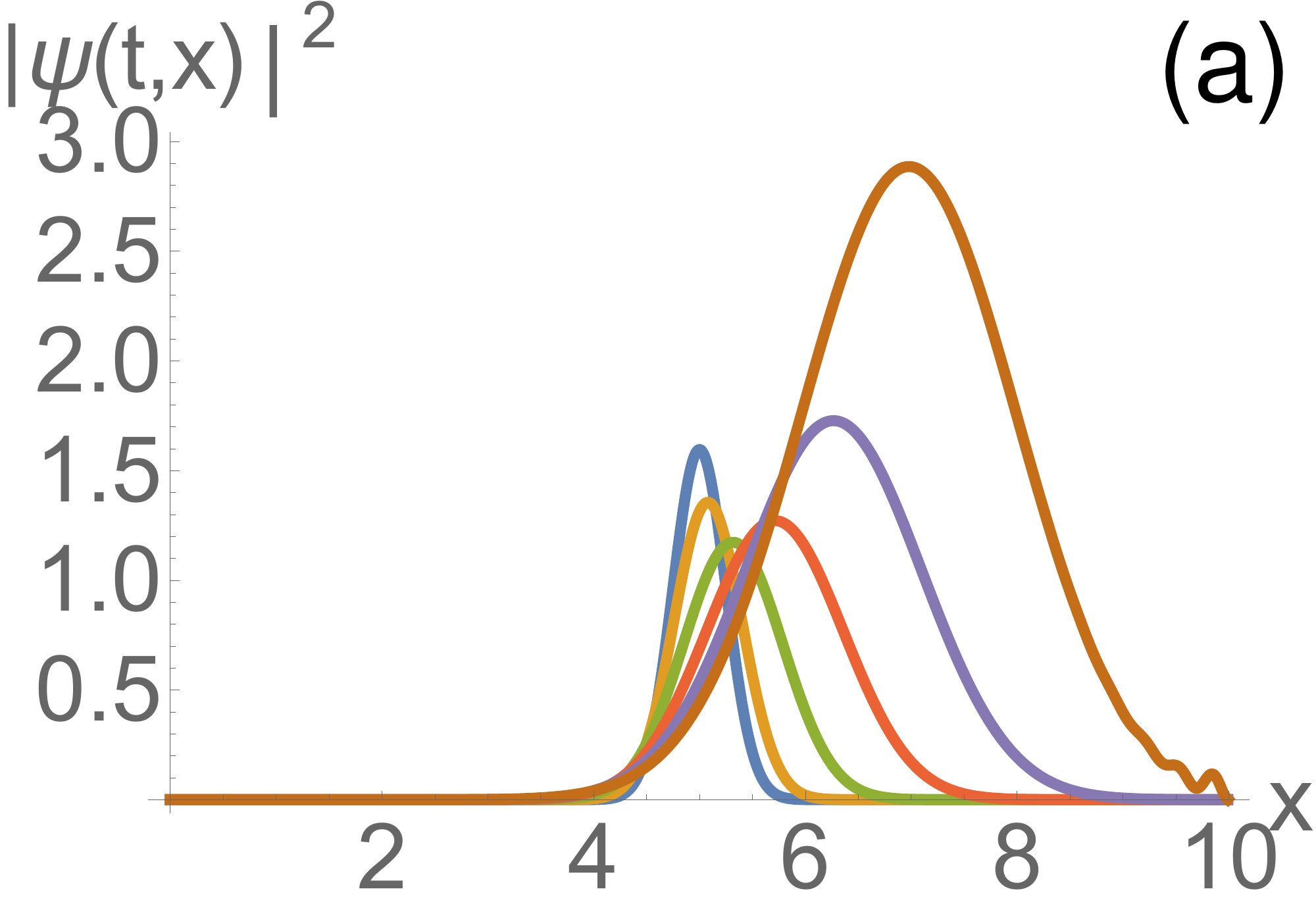}
    \includegraphics[width=0.45\columnwidth]{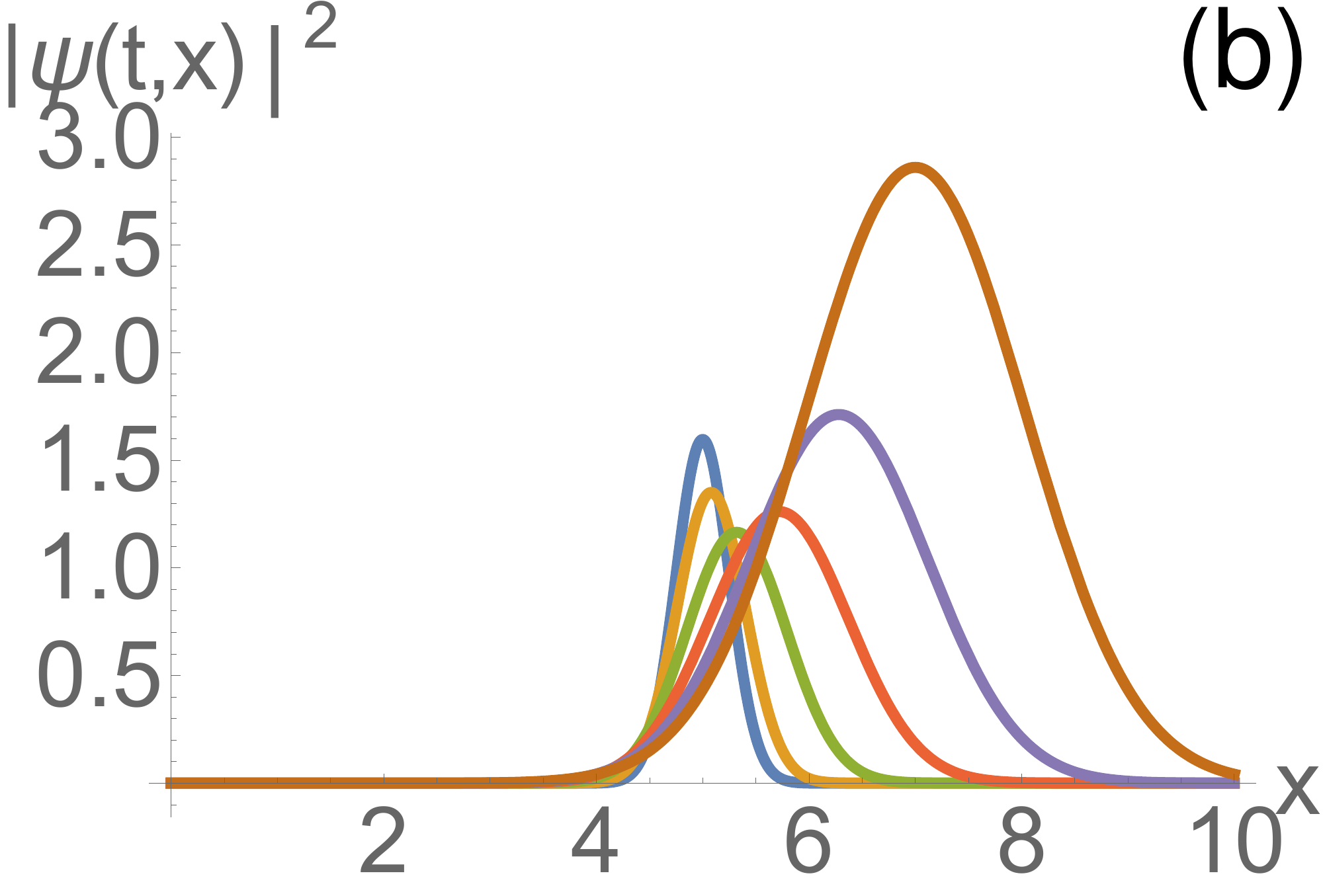}
    \caption[]{The time evolution of the Gaussian wave packet before it reaches the boundary. The center of the initial Gaussian wave packet is located at $x=5$. (a) Numerical result of the probability densities at various times $t$. (b) Analytical result of the probability densities at various times $t$ calculated by Eq.~(\ref{eq: Analy_result}).}
    \label{fig: Wave_timeEvo}
\end{figure}

\emph{Analytical solution.---}The Hamiltonian $H$ can be transformed to a Hermitian Hamiltonian $\bar{H}$ using the following similarity transformation 
\begin{align}
    \bar{H} = S^{-1} H S,
\end{align}
where $S$ is a multiplication operator satisfying $S(\psi)(x) = e^{b m x} \psi(x)$ and $\bar{H} = -\frac{\nabla^2}{2m}$ is the Hermitian Hamiltonian of the free particle. The time evolution can be calculated by using
\begin{align}
    \bra{x} e^{-i H t} \ket{\psi} = e^{b m x} \bra{x} e^{-i \bar{H} t} \ket{S^{-1} \psi}. 
    \label{eq: byHerm}
\end{align}
Note that $\ket{S^{-1} \psi}$ is also a Gaussian wave packet but with a constant $-2 b m \sigma^2$ center shift. Hence, $e^{-i \bar{H} t} \ket{S^{-1} \psi}$ describes the time evolution of the Gaussian wave packet under Hermitian dynamics, and has the following form,
\begin{align}
    |\bra{x} e^{-i \bar{H} t} \ket{S^{-1} \psi}|^2 \propto e^{-\frac{(x+2 b m \sigma^2)^2}{2 \sigma(t)^2}},
\end{align}
where $\sigma(t)^2 = \sigma^2 + \frac{t^2}{4 \sigma^2 m^2}$ describes the wave packet spreading under Hermitian dynamics \cite{supplemental}, i.e., the standard deviation of the wave packet grows with time. After the final multiplication of the exponential factor $e^{2b m x}$ by Eq.~(\ref{eq: byHerm}), the probability density is 
\begin{align}
    | \bra{x} e^{-i H t} \ket{\psi} |^2 = \frac{A}{\sqrt{2 \pi \sigma(t)^2}} e^{-\frac{\left[ x - 2 b m (\sigma(t)^2 - \sigma^2) \right]^2}{2 \sigma(t)^2}},
    \label{eq: Analy_result}
\end{align}
where $A = \exp\left\{ 2 b^2 m^2 \left[\sigma(t)^2-\sigma^2 \right] \right\}$, and the center (peak) of the wave packet $x_p$ relative to the starting point becomes time dependent, i.e., $x_p(t) = 2 b m \left[ \sigma(t)^2 - \sigma^2 \right]$. It is found that the Gaussian wave packet is accelerated and amplified during its time evolution. The probability densities as functions of $x$ at various times $t$ are shown in Fig.~\ref{fig: Wave_timeEvo}, which are snapshots of Fig.~\ref{fig: x-t_density}(a).

The analytical result of the wave packet trajectory is compared to the numerical result in Fig.~\ref{fig: x-t_density}(a).

\emph{Inelastic scattering.---}As shown in Figs.~\ref{fig: x-t_density}(b) and \ref{fig: x-t_density}(c), in this non-Hermitian model, the velocity of the incident wave packet $v_{\textrm{in}}$ does not equal the velocity of the reflected wave packet $v_{\textrm{ref}}$. This unconventional reflection is analogous to the inelastic scattering process in classical mechanics.

Let us provide a simple mechanism of this scattering process at the boundary. As above, the time evolution can be expressed as Eq.~(\ref{eq: byHerm}). Apart from the exponential factor $e^{b m x}$, the time evolution is described by the Hermitian dynamics $e^{-i \bar{H} t} \ket{S^{-1} \psi}$. In the Hermitian system, the wave packet with the initial speed $v_0$ will be reflected by the boundary with the speed $-v_0$. Because of the exponential factor $e^{b m x}$, the wave packet acquires an additional velocity
\begin{align}
    v_p(t) = \frac{d x_p(t)}{dt} = \frac{b t}{m \sigma^2 }
\end{align}
both before it reaches the boundary and after it reaches the boundary, and the time origin is the moment the wave packet begins to move.
Hence, the incident velocity is 
\begin{align}
    v_{\textrm{in}}(t) = v_0  + v_p(t) = v_0 + \frac{b t}{m \sigma^2 },
    \label{eq: v_in}
\end{align}
and the velocity of the reflected wave packet is 
\begin{align}
    v_{\textrm{ref}}(t) = -v_0 + v_p(t) = - v_0 + \frac{b t}{m \sigma^2 }.
    \label{eq: v_out}
\end{align}
A rigorous derivation of Eqs.~(\ref{eq: v_in}) and (\ref{eq: v_out}) is provided in SM \cite{supplemental}. Thus for a reflection process in this model, in general, $v_{\textrm{in}}(t_1) \neq - v_{\textrm{ref}}(t_2)$ ($t_1$ and $t_2$ are two times that before and after the reflection), which explains the inelastic scattering in this non-Hermitian model. 

It is worth noting that, when considering the reflection process at the right boundary, if $v_{\textrm{ref}}(t_2) \geqslant 0$, the peak of the wave packet is stuck at the boundary. Figure~\ref{fig: x-t_density}(a) shows such a case. For the reflection process at the left boundary, it can be seen from Eqs.~(\ref{eq: v_in}) and (\ref{eq: v_out}) that $|v_{\textrm{in}}(t_1)|<|v_{\textrm{ref}}(t_2)|$, as shown in Fig.~\ref{fig: x-t_density}(b), and the wave packet is decelerated before it hits the boundary. For the reflection process at the right boundary, it can be seen from Eqs.~(\ref{eq: v_in}) and (\ref{eq: v_out}) that $|v_{\textrm{in}}(t_1)|>|v_{\textrm{ref}}(t_2)|$, as shown in Fig.~\ref{fig: x-t_density}(c), and the wave packet is accelerated before it hits the boundary.

\begin{figure} 
    \centering 
    \includegraphics[width=0.48\columnwidth]{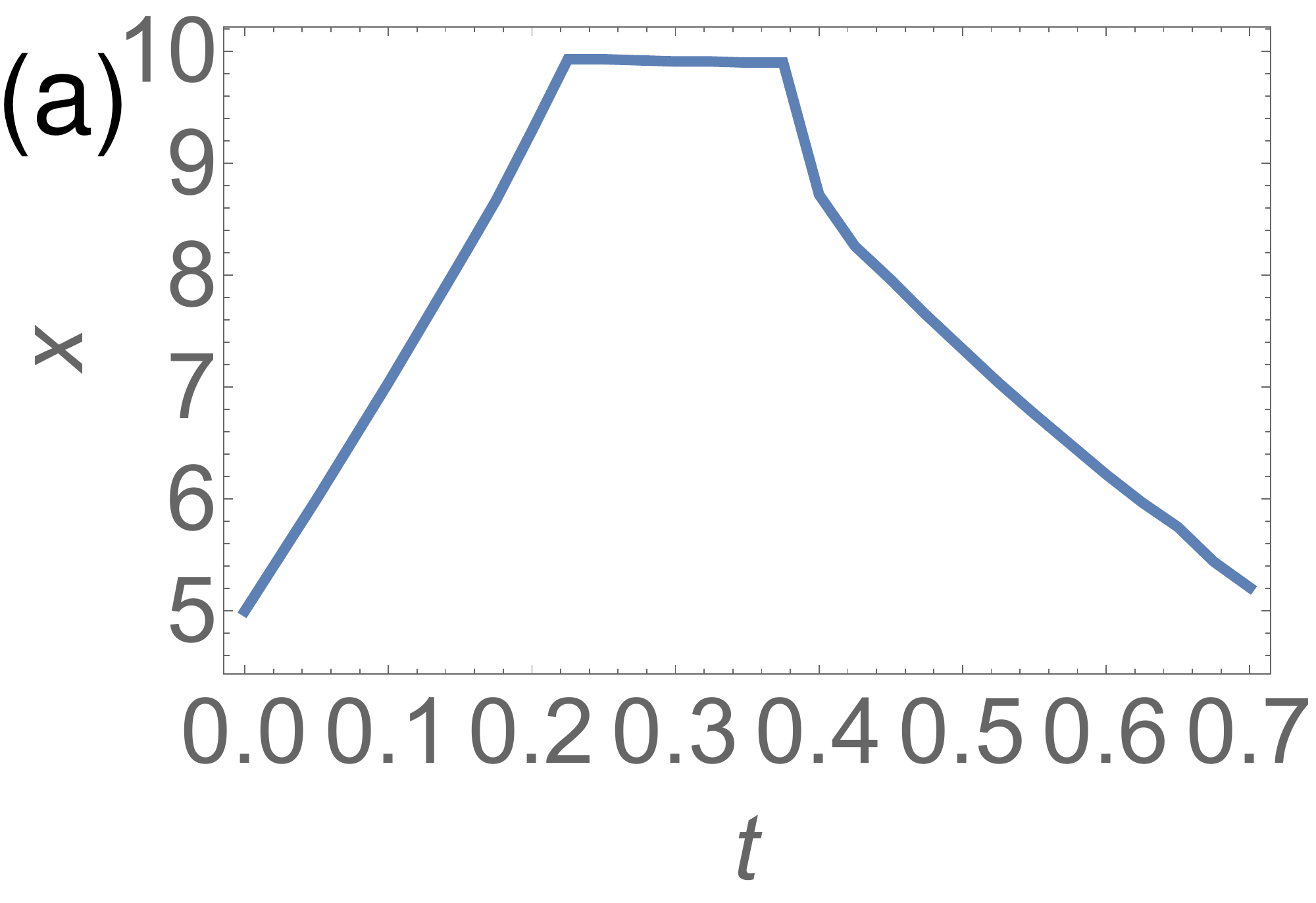}
    \includegraphics[width=0.48\columnwidth]{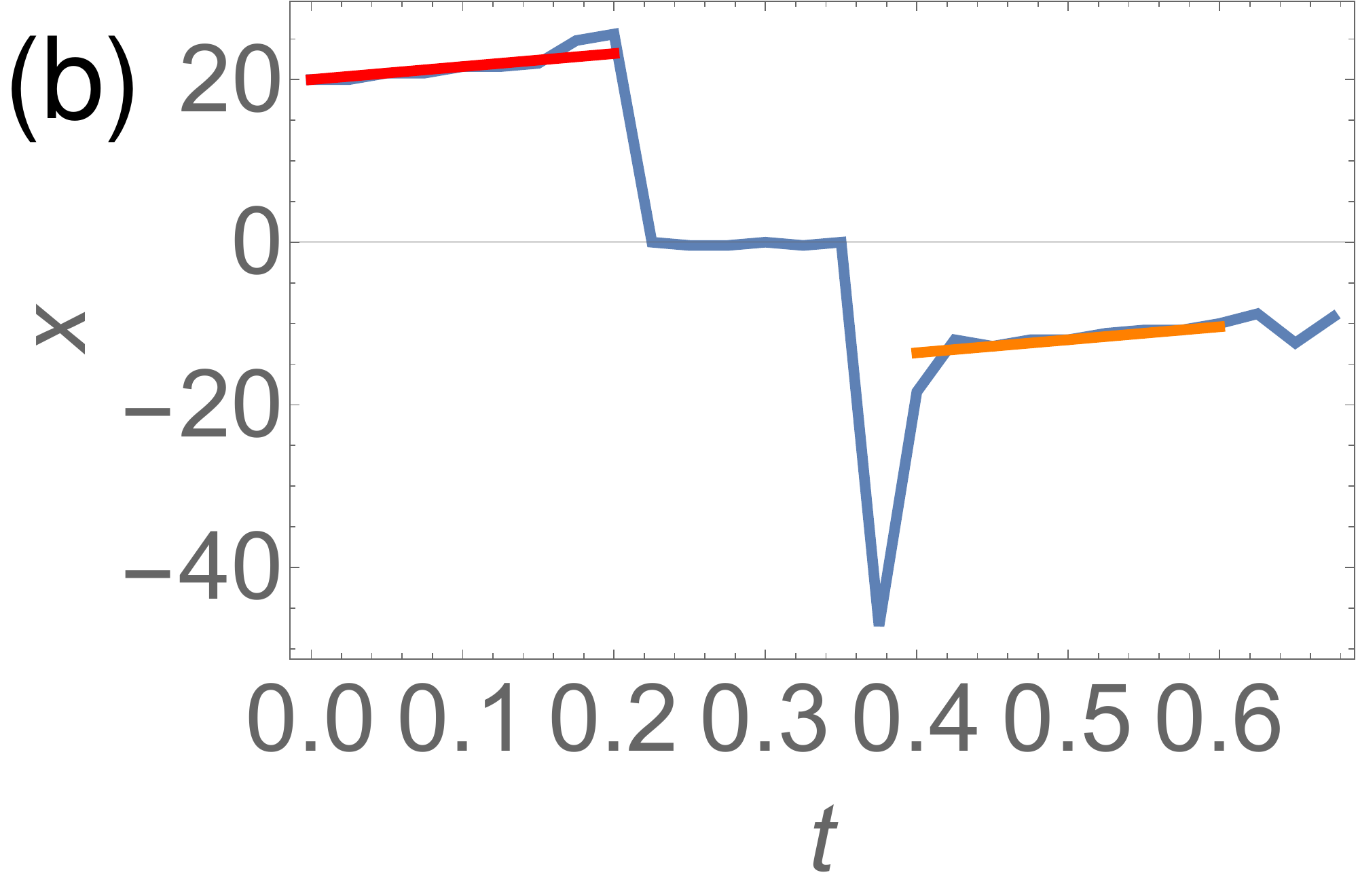}
    \caption[]{The trajectory and the speed of the wave packet as functions of $t$. (a) The position $x$ of the wave packet peak as a function of $t$. (b) The velocity $v$ of the wave packet peak as a function of $t$. The red line is the analytical result of $v_{\textrm{in}}(t)$ and the orange line is the analytical result of $v_{\textrm{ref}}(t)$ \cite{nonWavePacket_note}.}
    \label{fig: xv-t}
\end{figure}

To further verify our analytical result, we use the numerical data in Fig.~\ref{fig: x-t_density}(c) to generate the $x(t)$ and $v(t)$ relations, which are illustrated in Fig.~\ref{fig: xv-t}. Here, $x$ is the peak of the wave packet (position where the probability density is maximum), and $v(t)$ is the numerical differential of $x(t)$.
The numerical results fit well with the analytical results as shown in Fig.~\ref{fig: xv-t}(b). 

\emph{General result.---}The above findings can be generalized to other models. From the above exact result of the additional moving velocity of the HN model, we find that the wave packets spreading phenomenon is directly related to the additional wave packet motion. We want to mention that the wave packet spreading phenomenon is caused by the time evolution operator $e^{-i \bar{H} t}$, where $\bar{H}$ is a Hermitian operator, and is thus a result of Hermitian dynamics. 

For general non-Hermitian models with a uniform skin effect, assuming that $r=|\beta|$ is the exponential factor of the non-Hermitian skin modes, the time evolution can be expressed as
\begin{align}
    \bra{x} e^{-i H t} \ket{\psi} = r^x \bra{x} e^{-i \bar{H} t} \ket{S^{-1} \psi}.
\end{align}
The probability distribution of $e^{-i \bar{H} t} \ket{S^{-1} \psi}$ is Gaussian,
\begin{align}
    |\bra{x} e^{-i \bar{H} t} \ket{S^{-1} \psi}|^2 \propto e^{-\frac{(x+2\ln(r)\sigma^2)^2}{2 \sigma(t)^2}}.
\end{align}
Hence, the time evolution of the Gaussian wave packet with zero initial speed is approximately the following form, 
\begin{align}
    |\bra{x} e^{-i H t} \ket{\psi}|^2 \approx  c \, r^{2x} e^{-\frac{(x+2\ln(r)\sigma^2)^2}{2 \sigma(t)^2}}. 
    \label{eq: genTimeEvo}
\end{align}
Before the wave packet reaches the boundary, the wave packet peak satisfying $\frac{d}{dx} \vert_{x_p} |\bra{x} e^{-i H t} \ket{\psi}|^2 =0$ is 
\begin{align}
    x_p(t) = 2 \ln(r) \left[ \sigma(t)^2 - \sigma(0)^2 \right]
    \label{eq: xp_sigma}
\end{align}
and the velocity of the wave packet peak is 
\begin{align}
    v_p(t) = 2 \ln(r) \frac{d \, \sigma(t)^2}{d t}.
\end{align}
To make $x_p(t) \neq 0$, $r$ must not be 1 and $\sigma(t)$ must change over time, and the additional motion of the wave packet depends on both the magnitude of NHSE and the speed of wave packet spreading. If the wave packet has the initial momentum $k_0$, then the incident velocity and reflected velocity of the wave packet are  
\begin{align}
    v_{\textrm{in}}(t) & = \frac{\partial E(k)}{\partial k}\vert_{k_0}  + v_p(t) , \notag \\
    v_{\textrm{ref}}(t) & = \frac{\partial E(k)}{\partial k}\vert_{k_1} + v_p(t) ,
\end{align}
where $E(k)$ is the energy dispersion of the wave packet mode, and $k_1$ satisfying $E(k_1) = E(k_0)$ is the central momentum of the reflected wave packet. The existence and the identical sign of the $v_p(t)$ terms in $v_{\textrm{in}}(t)$ and $v_{\textrm{ref}}(t)$ cause the inelastic scattering, and it can be seen that the dynamic skin effect can still happen even if non-Hermiticity is only turned on near the boundary (see Sec.~IX of the SM \cite{supplemental}).
The above formulas are further verified in the non-Hermitian Su-Schrieffer-Heeger (SSH) model. 

Consider the non-Hermitian SSH model \cite{yao2018, PhysRevLett.116.133903} with the Bloch Hamiltonian
\begin{align}
    H(k) = \left( t_1 + t_2 \cos k \right) \sigma_x + \left( t_2 \sin k + i \frac{\gamma}{2} \right) \sigma_y,
\end{align}
where $\sigma_{x,y,z}$ are the Pauli matrices, with $\sigma_z = 1(-1)$ corresponding to the $A(B)$ sublattice. The exponential factor of the non-Hermitian skin modes is $r = \sqrt{|\frac{t_1 - \gamma/2}{t_1 + \gamma/2 }|}$. As discussed in Ref.~\cite{yao2018}, by taking $S=\textrm{diag}(1,r,r,r^2,r^2, \dots, r^{L-1},r^{L-1},r^L)$, $\bar{H} = S^{-1} H S$ is a Hermitian Hamiltonian with $k$-space form 
\begin{align}
    \bar{H}(k) = \left( \bar{t}_1 + t_2 \cos k \right) \sigma_x + t_2 \sin k \, \sigma_y ,
    \label{eq: HermH}
\end{align}
where $\bar{t}_1 = \sqrt{(t_1-\gamma/2)(t_1 + \gamma/2)}$. In the following, we numerically verify the additional motion formula (\ref{eq: xp_sigma}) in this model. Instead of deriving an analytical formula of $\sigma(t)^2$ for this model, we use numerical data to determine $\sigma(t)$. Our objective is to examine the relation between the wave packet spreading and the additional motion, not to find an exact formula for $\sigma(t)$.

For a Gaussian wave packet, there is a relation between the half-wave width $\delta$ and the standard deviation $\sigma$, which is 
\begin{align}
    \frac{\delta}{2} = \sqrt{2 \ln(2)} \sigma. 
    \label{eq: halfW_sigma}
\end{align}
The half-wave width is easy to obtain from the numerical data, hence, the standard deviation can be calculated by using Eq.~(\ref{eq: halfW_sigma}).

In our simulation, we set $t_1=2$, $t_2=1$, and $N=500$ for the number of unit cells. The initial state is
\begin{align}
    \braket{x A | \psi(0)} & = \frac{1}{(2 \pi \sigma^2)^{\frac{1}{4}}} e^{-\frac{x^2}{4 \sigma^2} + i k_0 x} , \notag \\
    \braket{x B | \psi(0)} & = 0,
\end{align}
where $A,B$ are two sites in the unit cell and $\sigma = 20$. The initial velocity $v_{0,\pm} = \frac{\partial E_{\pm}(k)}{\partial k} \vert_{k_0}$, where $E_{\pm}(k) = \pm \sqrt{\left( \bar{t}_1 + t_2 \cos k \right)^2 + \left( t_2 \sin k  \right)^2}$ are two energy levels of the Hermitian Hamiltonian (\ref{eq: HermH}). For $k_0=0$, $v_{0,\pm} = 0$. For $k_0 \neq 0$, in general, $v_{0,+} = -v_{0,-} \neq 0$, there are two modes moving in different directions, and their velocities are 
\begin{align}
    v_{+}(t) = v_{0,+} + v_p(t) , \notag \\
    v_{-}(t) = v_{0,-} + v_p(t).
\end{align}
For $k_0 =0$, due to $v_{0,\pm}=0$, the trajectory of these two modes coincides, and we only need to consider one of them. In the following, let us first consider the $k_0 = 0$ case.

\begin{figure} 
    \includegraphics[width=0.52\columnwidth]{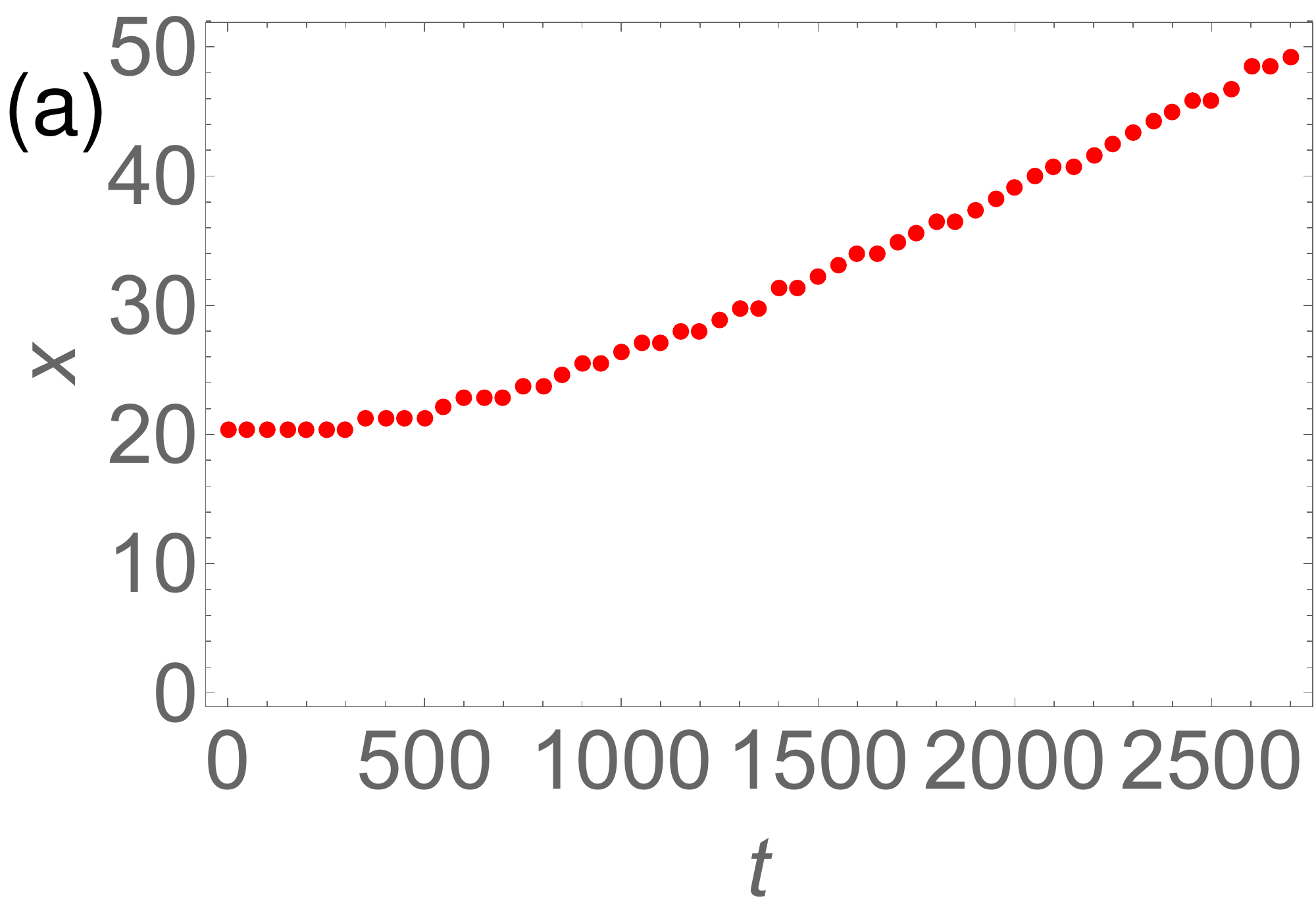}
    \includegraphics[width=0.40\columnwidth]{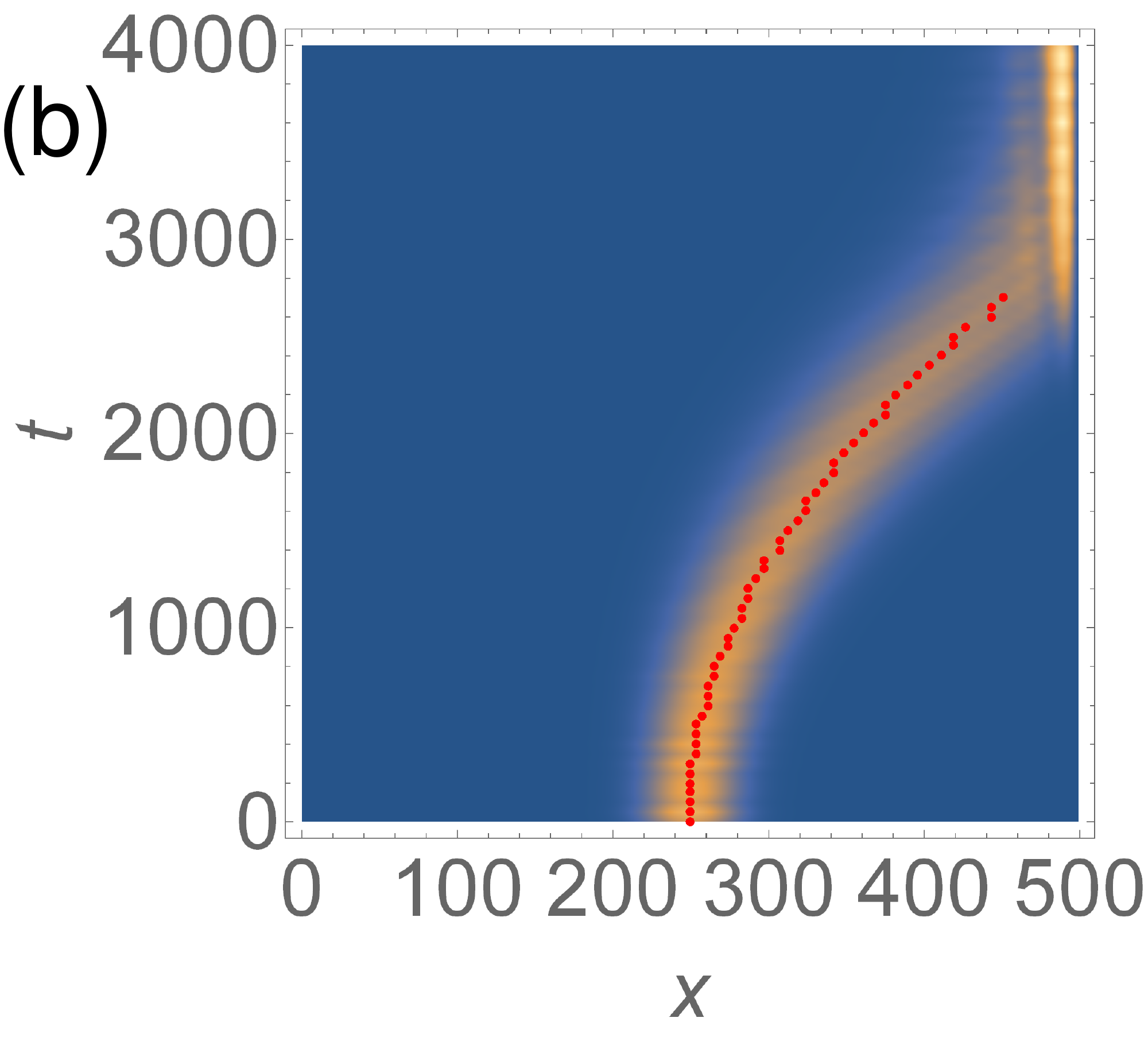}
    \caption[]{Wave packet acceleration and inelastic scattering in the non-Hermitian SSH model. (a) $\sigma(t)$ as a function of $t$, which describes the wave packet spreading. (b) Numerical simulation of the probability distribution $|\braket{x | \psi(t)}|^2$ (normalized) with $k_0=0$, shown as a function of $x$ and $t$. The wave packet is accelerated before it reaches the boundary. The wave packet trajectory calculated by using $\sigma(t)$ and Eq.~(\ref{eq: xp_sigma}) is shown as red points. After the wave packet reaches the boundary, it remains there and is not reflected.}
    \label{fig: SSH_density}
\end{figure}

Let $\gamma=-0.2$, so the standard deviation $\sigma(t)$ is plotted in Fig.~\ref{fig: SSH_density}(a) as a function of $t$, and the probability distribution $|\psi(t, x)|^2 = |\braket{x A | \psi(t)}|^2 + |\braket{x B | \psi(t)}|^2$ is plotted as a function of $t$ and $x$ in Fig.~\ref{fig: SSH_density}(b). As shown in Fig.~\ref{fig: SSH_density}(b), Eq.~(\ref{eq: xp_sigma}) is quite accurate.

\begin{figure} 
    \centering 
    \includegraphics[width=0.91\columnwidth]{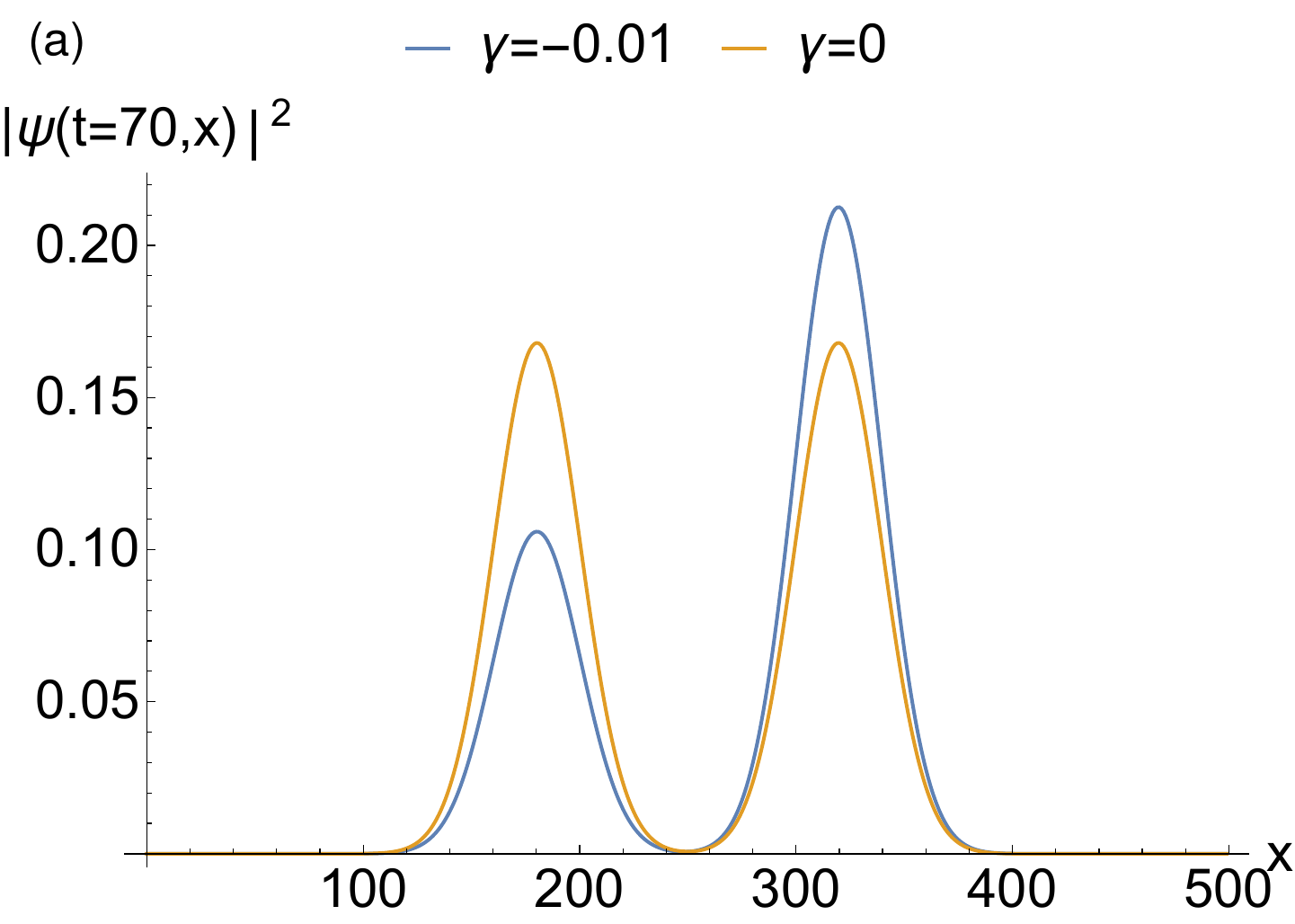}
    \includegraphics[width=0.45\columnwidth]{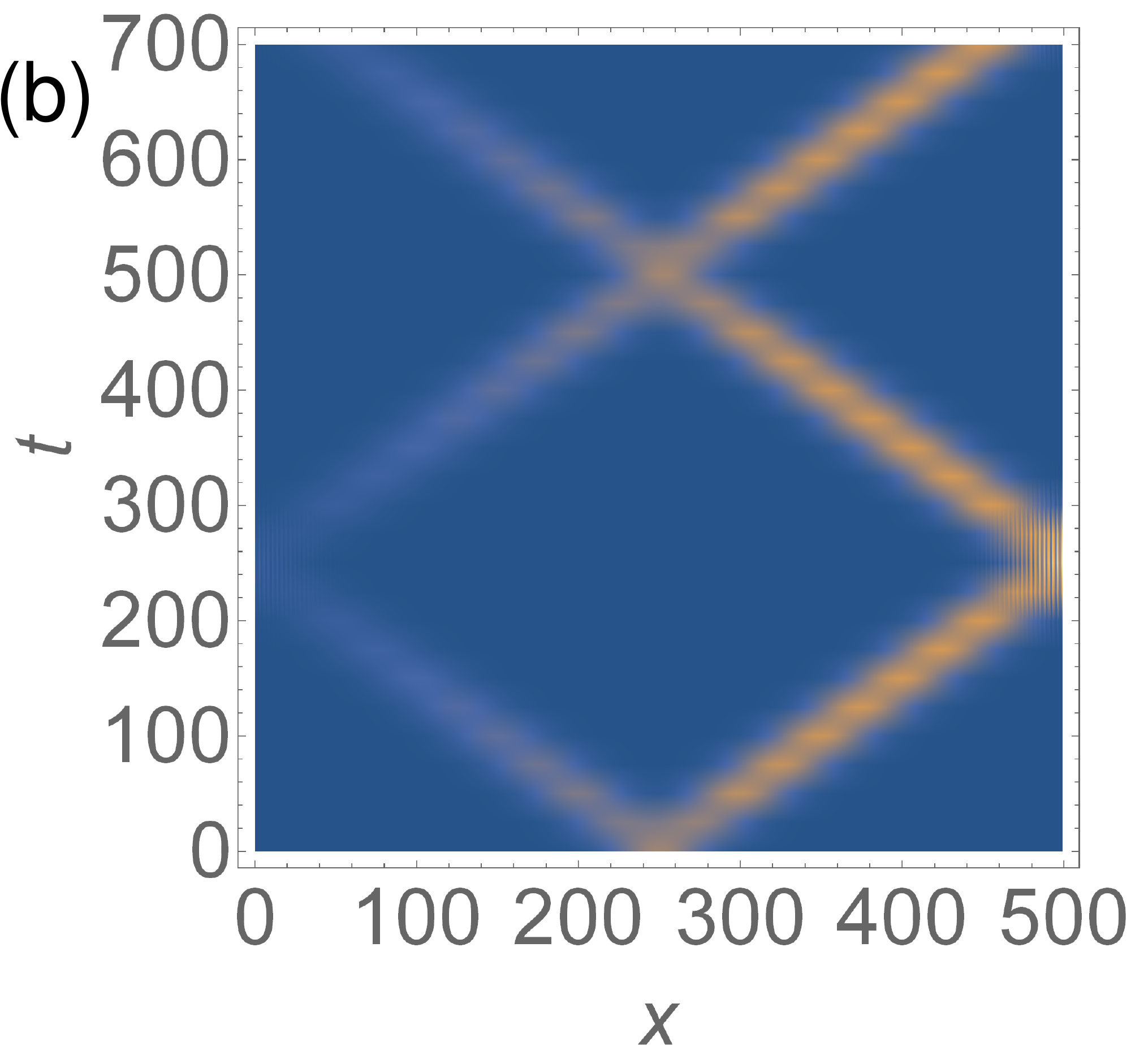}
    \includegraphics[width=0.45\columnwidth]{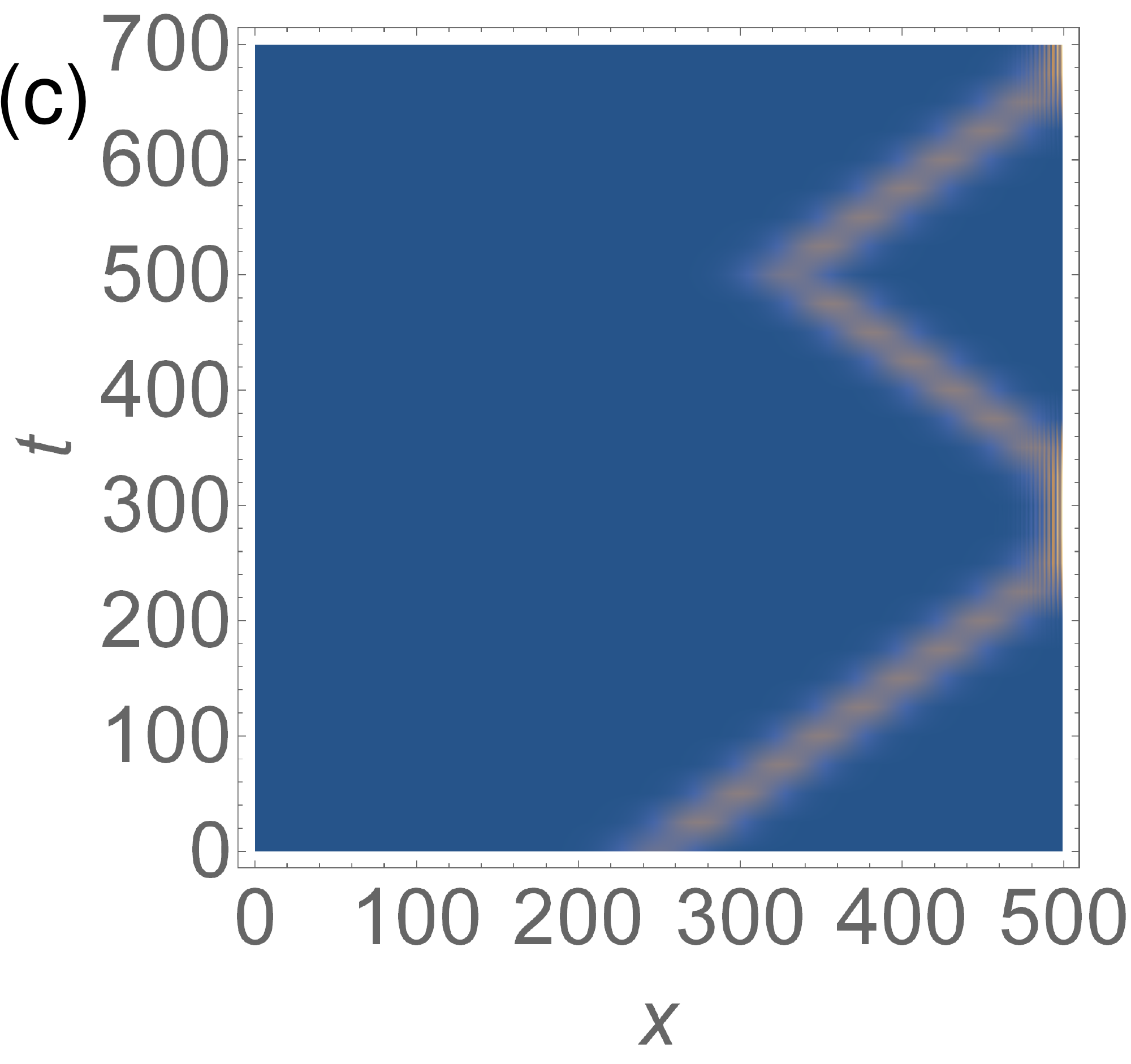}
    \caption[]{(a) The probability distributions of the Hermitian case ($\gamma=0$) and non-Hermitian case ($\gamma=-0.01$) with $k_0 =2$ at $t=70$. (b), (c) Numerical simulation of the probability distribution $|\braket{x | \psi(t)}|^2$ (normalized) as a function of $x$ and $t$, with $k_0 =2$. (b) When $\gamma=-0.01$, two distinct modes traveling in opposite directions can be seen. Here, the non-Hermitian parameter $\gamma$ is small, therefore the effects of wave packet acceleration and inelastic scattering are negligible. (c) When $\gamma=-0.2$, we can only see the trajectory of the higher wave packet peak mode. In the first reflection process, the velocity of the incident wave packet $|v_{\textrm{in}}|$ is larger than the velocity of the reflected wave packet $|v_{\textrm{ref}}|$.}
    \label{eq: twoModes}
\end{figure}

Now let us consider the $k_0 \neq 0$ case. Figures \ref{eq: twoModes}(a) and \ref{eq: twoModes}(b) show there are two modes traveling in opposite directions. In the non-Hermitian case, due to the exponential factor $r^{2x}$ in Eq.~(\ref{eq: genTimeEvo}), one of the wave packet peaks is lower than the other, and the wave packet with the lower peak is always on the left for $\gamma<0$. For a big enough $r$, the lower wave packet peak mode can be unrecognizable, as seen in Fig.~\ref{eq: twoModes}(c), and we can only see the trajectory of the higher wave packet peak mode. However, the wave packet with a lower peak still exists, and when the wave packet with a lower peak and the wave packet with a higher peak meet, they will pass through each other, and the wave packet with the lower peak will become the one with the higher peak. Specifically, the fold line pattern at $(x,t) \approx (330,500)$ in Fig.~\ref{eq: twoModes}(c) is not due to the deceleration effect; rather, it is formed by the wave packet with a lower peak coming from the left and becoming the wave packet with a higher peak.
Snapshots of a similar wave meeting process are shown in SM \cite{supplemental}.

\emph{Conclusions.---}In this Letter, we study the dynamics of non-Hermitian systems under the OBC, confirm that the dynamic skin effects exist robustly, and unveil the mechanism of the formation of the dynamic skin effect. We find that the Gaussian wave packet can be accelerated and amplified in non-Hermitian systems, and this additional acceleration motion of the wave packet is found to be the reason for the dynamic skin effect. The wave packet acceleration and inelastic scattering are explained by the interplay of the NHSE and the Hermitian wave packet spreading. A complete analytical result is established in the HN model and a universal formula for systems with a uniform skin effect is constructed. Furthermore, as an unconventional reflection process, the dynamic skin effect still occurs when one only turns on non-Hermiticity close to the boundary. The physics can be generalized to other non-Hermitian systems, which will be left for future study. Our work paves the way for studying dynamic skin effects of non-Hermitian systems.

{\it Acknowledgements.---}This work is supported by NSFC under Grant No.11275180.

\bibliography{main}

\begin{thebibliography}{55}
\bibitem [{nor()}]{normalization_note}%
  \BibitemOpen
  \href@noop {} {}\bibinfo {howpublished} {To clearly show the wave packet
  trajectory, the probability density $|\braket{x | \psi(t)}|^2$ has been
  normalized for each $t \neq 0$, and this normalization does not affect the
  trajectory of the wave packet peak and the shape of the wave
  packet.}\BibitemShut {Stop}%
\end{thebibliography}

\end{document}


\newcommand{\ii}{\text{i}}
\newcommand{\U}{U}
\newcommand{\V}{V}
\newcommand{\BZ}{\left[ 0, 2\pi \right]}
\newcommand{\CZ}{\left[ 0, 1 \right]}

\title{Supplemental Material for ``Dynamic skin effects in non-Hermitian systems''}

\author{Haoshu Li}
    \email{lihaoshu@mail.ustc.edu.cn}
    \affiliation{Department of Modern Physics, University of Science and Technology of China, Hefei 230026, China}

\author{Shaolong Wan}
    \email{slwan@ustc.edu.cn}

    \affiliation{Department of Modern Physics, University of Science and Technology of China, Hefei 230026, China}

\maketitle
\onecolumngrid
\tableofcontents

\section{Non-unitary quantum walks}
The effective Schr{\"o}dinger equation with the non-Hermitian Hamiltonian can be physically realized by the non-unitary quantum walks \cite{PhysRevLett.119.130501, PhysRevA.98.063847, PhysRevLett.122.020501, PhysRevLett.123.230401, Xiao2020}. As in Ref.~\cite{PhysRevLett.119.130501, Xiao2020}, consider a non-unitary quantum walk governed by a non-unitary Floquet operator $U$. The state $\ket{\psi(t)} = U^t \ket{\psi(0)}$ is evolved at discrete time step $t$. This quantum walk simulates the non-unitary time evolution driven by a non-Hermitian effective Hamiltonian $H_{\text{eff}}$, with $U=e^{-i H_{\text{eff}}}$. It follows that the time-evolved state $\ket{\psi(t)}$ satisfies the Schr{\"o}dinger equation
\begin{align}
    i \partial_t \ket{\psi(t)} = H_{\text{eff}} \ket{\psi(t)}.
\end{align}
The norm of the wave function at the space position $x$ ($|\braket{x | \psi(t)}|$) is an experimental measurable quantity \cite{PhysRevLett.119.130501, PhysRevA.98.063847, Xiao2020} and is the main focus in this manuscript.

\section{Relation between the non-Hermitian dynamics and Hermitian dynamics}
\subsection{Similarity transformation}
Consider a non-Hermitian system which can be transformed to a Hermitian system by a similarity transformation. Assume the real-space single-particle Hamiltonian of the non-Hermitian system under the open boundary condition (OBC) is $H$, the similarity transformation is $S$, and the Hamiltonian of the Hermitian system is $\bar{H}$, by the previous assumption, 
\begin{align}
    \bar{H} = S^{-1} H S.
    \label{eq: sim}
\end{align}
Note that we require the Hermitian Hamiltonian $\bar{H}$ having the translation symmetry except at the boundary.

Assume a set of physical quantities $\psi_i$ that evolve over time and are governed by the effective non-Hermitian Hamiltonian $H$, i.e., the equation of motion is
\begin{align}
    i \partial_t{\ket{\bm{\psi}}} = H \ket{\bm{\psi}} ,
    \label{eq: EOM}
\end{align}
where $\ket{\bm{\psi}} = (\psi_1, \psi_2, \ldots)^{T}$.

By Eq.~(\ref{eq: sim}) and Eq.~(\ref{eq: EOM}), the initial state $\ket{\bm{\psi}}$ evolves as 
\begin{align}
    \bra{x} e^{-i H t} \ket{\bm{\psi}} = \bra{x} S e^{-i \bar{H} t} S^{-1} \ket{\bm{\psi}}.
    \label{eq: propagate}
\end{align}
The similarity transformation $S$ usually satisfies $S \ket{x} = r^x \ket{x}$, hence,
\begin{align}
    \bra{x} S e^{-i \bar{H} t} S^{-1} \ket{\bm{\psi}} = r^x \bra{x} e^{-i \bar{H} t} S^{-1} \ket{\bm{\psi}} .
    \label{eq: propagator}
\end{align}
By Eq.~(\ref{eq: propagate}) and Eq.~(\ref{eq: propagator}), it follows that 
\begin{equation}
    \bra{x} e^{-i H t} \ket{\bm{\psi}} = r^x \bra{x} e^{-i \bar{H} t} \ket{S^{-1}\bm{\psi}}.
    \label{eq: timeEvo}
\end{equation}
It implies that the time evolution of the initial state $\ket{\bm{\psi}}$ can be expressed by the Hermitian time evolution $e^{-i \bar{H} t} \ket{S^{-1}\bm{\psi}}$ and the exponential factor $r^{x}$. 

\subsection{Examples of similarity transformable non-Hermitian systems} \label{sec: HN_SSH}
There are many similarity transformable non-Hermitian systems, in this section, we discuss two of them which are the Hatano-Nelson (HN) model \cite{hn} and the non-Hermitian Su-Schrieffer-Heeger (SSH) model \cite{yao2018, PhysRevLett.116.133903}.

First, let us consider the HN model \cite{hn}
\begin{align}
    H = \sum_{n=1}^{L-1} (t_{-1} c^{\dagger}_{n+1} c_n + t_1 c^{\dagger}_n c_{n+1}).
\end{align}
Under the OBC, the real space Hamiltonian is 
\begin{align}
    H = \begin{pmatrix}
        0 & t_1 & 0 & \cdots & \cdots & 0 \\
        t_{-1} & 0 & t_1 & 0  & \cdots & 0 \\
        0 & t_{-1} & 0 & t_1 & \cdots & 0 \\
        \vdots & \ddots & \ddots & \ddots & \ddots & \vdots \\
        0 & \cdots & 0 & t_{-1} & 0 & t_1 \\
        0 & \cdots & \cdots & 0 & t_{-1} & 0
    \end{pmatrix}_{L \times L} .
\end{align}
Let $S = \textrm{diag}(r, r^2, \ldots, r^N)$, then 
\begin{align}
    \bar{H} & = S^{-1} H S \notag \\
    & = \begin{pmatrix}
        0 & t_1 r & 0 & \cdots & \cdots & 0 \\
        t_{-1} r^{-1} & 0 & t_1 r & 0  & \cdots & 0 \\
        0 & t_{-1} r^{-1} & 0 & t_1 r & \cdots & 0 \\
        \vdots & \ddots & \ddots & \ddots & \ddots & \vdots \\
        0 & \cdots & 0 & t_{-1} r^{-1} & 0 & t_1 r \\
        0 & \cdots & \cdots & 0 & t_{-1} r^{-1} & 0
    \end{pmatrix}_{L \times L} .
\end{align}
Taking $r=\sqrt{\frac{t_{-1}}{t_1}}$ for positive $t_1$ and $t_{-1}$, then $\bar{H}$ becomes a Hermitian Hamiltonian.

Second, let us consider the non-Hermitian SSH model whose Bloch Hamiltonian is 
\begin{align}
    H(k) = \left( t_1 + t_2 \cos k \right) \sigma_x + \left( t_2 \sin k + i \frac{\gamma}{2} \right) \sigma_y,
\end{align}
where $\sigma_{x,y,z}$ are the Pauli matrices, with $\sigma_z = 1(-1)$ corresponding to A(B) sublattice. As discussed in Ref.~\cite{yao2018}, by taking $S=\textrm{diag}(1,r,r,r^2,r^2, \dots, r^{L-1},r^{L-1},r^L)$ and $r = \sqrt{|\frac{t_1 - \gamma/2}{t_1 + \gamma/2 }|}$, $\bar{H} = S^{-1} H S$ is a Hermitian Hamiltonian with k-space form 
\begin{align}
    \bar{H}(k) = \left( \bar{t}_1 + t_2 \cos k \right) \sigma_x + t_2 \sin k \, \sigma_y ,
\end{align}
where $\bar{t}_1 = \sqrt{(t_1-\gamma/2)(t_1 + \gamma/2)}$.

\section{Generalized by spectral decomposition}
In general, a non-defective non-Hermitian Hamiltonian $H$ has the following spectral decomposition 
\begin{align}
    H = \sum_n E_n \ket{n R} \bra{n L},
\end{align}
where $\ket{n R}$ is the right eigenstate of $H$ with eigenenergy $E_n$ and $\ket{n L}$ is the corresponding left eigenstate.

The time evolution operator of $H$ is  
\begin{align}
    e^{-i H t} = & \sum_{m = 0}^{+\infty} \frac{(-i H t)^m}{m!} \notag \\
    = & \sum_{m = 0}^{+\infty} \sum_{n_1,\ldots,n_m} \frac{(-i E_{n_1} t) (-i E_{n_2} t) \cdots (-i E_{n_m} t)}{m!} \ket{n_1 R} \braket{n_1 L| n_2 R} \cdots \ket{n_m L} \notag \\
    = & \sum_{m = 0}^{+\infty} \sum_n \frac{(-i E_n t)^m}{m!} \ket{n R} \bra{n L} \notag \\
    = & \sum_n e^{-i E_n t} \ket{n R} \bra{n L},
\end{align}
where in the third equality the biorthogonal relation $\braket{n_i L|n_j R} = \delta_{n_i n_j}$ has been used.

We can use wave vector $\beta$ in the GBZ to label the bulk energy eigenstate $\ket{n R}$ and we simply denote it by $\ket{\beta R}$. In any system without topological edge states, any energy eigenstate $\ket{n R}$ corresponds to a $\ket{\beta R}$. At least in the bulk, $\ket{\beta R}$ and $\ket{\beta L}$ has the following form 
\begin{align}
    \braket{x | \beta R} = |\beta|^x \phi_{\beta}(x), \notag \\
    \braket{x | \beta L} = |\beta|^{-x} \varphi _{\beta}(x),
\end{align}
where $\phi_{\beta}(x)$ and $\varphi _{\beta}(x)$ are non-exponential functions (they are usually linear combinations of trigonometric functions). 

Hence, the time evolution of $\ket{\psi}$ is given by
\begin{align}
    \bra{x} e^{-i H t} \ket{\psi} = \sum_{\beta} |\beta |^{x} \phi_{\beta}(x) e^{-i E(\beta) t} \braket{\beta L| \psi}.
    \label{eq: Prop}
\end{align}
Eq.~(\ref{eq: Prop}) is a generalization of Eq.~(\ref{eq: timeEvo}); instead of the appearance of the single exponential factor $r^{x}$ in Eq.~(\ref{eq: timeEvo}), Eq.~(\ref{eq: Prop}) contains all of the exponential factors' contributions. If all skin modes have the same spatial exponential factor $r=|\beta|$, Eq.~(\ref{eq: Prop}) can be written as 
\begin{align}
    \bra{x} e^{-i H t} \ket{\psi} = r^{x}\sum_{\beta} \phi_{\beta}(x) e^{-i E(\beta) t} \braket{\beta L| \psi}.
\end{align}
If the spatial exponential factors of all skin modes are roughly the same, i.e., $|\beta| \approx r > 1$ or $|\beta| \approx r < 1$, then the mean-field treatment provides a good approximation
\begin{align}
    \bra{x} e^{-i H t} \ket{\psi} & \approx \sum_{\beta} \braket{|\beta |}^x \phi_{\beta}(x) e^{-i E(\beta) t} \braket{\beta L| \psi}, \notag \\
    & = r^{x}\sum_{\beta} \phi_{\beta}(x) e^{-i E(\beta) t} \braket{\beta L| \psi},
\end{align}
where the energy dispersion $E(\beta)$ causes the wave spreading phenomenon. Hence, if the initial state $\ket{\psi}$ is the Gaussian wave packet with $\braket{x | \psi} = \frac{1}{(2 \pi \sigma^2)^{\frac{1}{4}}} e^{-\frac{x^2}{4 \sigma^2}}$, its time evolution is 
\begin{align}
    |\bra{x} e^{-i H t} \ket{\psi}|^2 \approx c \, r^{2x} e^{-\frac{(x+2\ln(r)\sigma^2)^2}{2 \sigma(t)^2}}
\end{align}
since $|\sum_{\beta} \phi_{\beta}(x) e^{-i E(\beta) t} \braket{\beta L| \psi}|^2 \approx c \, e^{-\frac{(x+2\ln(r)\sigma^2)^2}{2 \sigma(t)^2}}$.

\section{Additional moving velocity}
\subsection{Why the additional motion of the wave packet happens}
In this section, we show that wave packets have an additional moving velocity in non-Hermitian systems. Consider the Hamiltonian we discussed in the main text 
\begin{align}
    H = -\frac{\nabla^2}{2m} + b \nabla - \frac{b^2 m}{2},
    \label{eq: conHN}
\end{align}
where $\nabla = \frac{d}{dx}$ in one-dimensional systems and $b \in \mathbb{R}$.
The boundary condition is taken to be the Dirichlet boundary condition 
\begin{align}
    \psi |_{\partial \Omega} = 0,
\end{align}
which corresponds to the system in a one-dimensional infinitely deep square potential well. The Hamiltonian (\ref{eq: conHN}) can be expressed in a simple form 
\begin{align}
    H = -\frac{1}{2 m} (\frac{d}{d x} - b m)^2.
\end{align}
Let $S$ be the following multiplication operator acting on the wave function 
\begin{align}
    S(\psi)(x) = e^{b m x} \psi(x) .
\end{align}
By applying $S$ an exponential factor $e^{b m x}$ is multiplied on $\psi(x)$ and the Dirichlet boundary condition is preserved by $S$. Let 
\begin{align}
    \bar{H} = S^{-1} H S.
\end{align}
Due to $(\frac{d}{d x} - b m) S = S \frac{d}{d x}$, it follows that 
\begin{align}
    \bar{H} = -\frac{1}{2 m} (\frac{d}{d x})^2 = -\frac{\nabla^2}{2m},
\end{align}
where $\bar{H}$ is a Hermitian Hamiltonian.

The time evolution equation (\ref{eq: timeEvo}) away from the boundary (transforming to k-space is valid) can be written as
\begin{align}
    \bra{x} e^{-i H t} \ket{\psi} & = \int dk \, dk^{\prime} \, r^{x} \braket{x | k^{\prime}} \bra{k^{\prime}} e^{-i \bar{H} t} \ket{k} \braket{k | S^{-1} \psi} \notag \\
    & = r^{x} \int d k \, \braket{x | k}  e^{-i \bar{H}(k) t} \braket{k | S^{-1} \psi},
    \label{eq: phiT}
\end{align}
where $r=e^{b m}$. Note that apart from the $r^x$ exponential factor, the above equation is about the Hermitian dynamics $e^{-i \bar{H} t} \ket{S^{-1} \psi}$.

Consider the Gaussian wave packet 
\begin{align}
    \braket{x | \psi} = \frac{1}{(2 \pi \sigma^2)^{\frac{1}{4}}} e^{-\frac{x^2}{4 \sigma^2}}. 
\end{align}
It is a normalized wave function, i.e., $\braket{\psi | \psi} = 1$. Furthermore, $\ket{S^{-1} \psi}$ is also a Gaussian wave packet since
\begin{align}
    \braket{x | S^{-1} \psi} & = e^{-bm x} \frac{1}{(2 \pi \sigma^2)^{\frac{1}{4}}} e^{-\frac{x^2}{4 \sigma^2}} \notag \\
    & = \frac{1}{(2 \pi \sigma^2)^{\frac{1}{4}}} e^{-\frac{(x+2 bm \sigma^2)^2}{4 \sigma^2}} e^{b^2 m^2 \sigma^2}.
\end{align}
Compared to the initial wave packet, its center has a constant shift.

By Eq.~(\ref{eq: phiT}), 
\begin{align}
    \bra{x} e^{-i H t} \ket{\psi} = \frac{\sqrt[4]{\frac{2}{\pi }} \exp \left(b^2 m^2 \sigma ^2-\frac{m \left(2 b m \sigma ^2+x\right)^2}{4 m \sigma ^2+2 i t}\right)}{\sqrt{2 \sigma +\frac{i t}{m \sigma }}} e^{b m x}
\end{align}
and 
\begin{align}
    |\bra{x} e^{-i H t} \ket{\psi}|^2 & = c \, e^{2 b^2 m^2 \sigma^2} e^{2 b m x} e^{-\frac{(x+2b m \sigma^2)^2}{2 \sigma(t)^2}} \notag \\
    & = c \, e^{2 b^2 m^2 \left[\sigma(t)^2 - \sigma^2\right]} e^{-\frac{\left[ x - 2 b m (\sigma(t)^2 - \sigma^2) \right]^2}{2 \sigma(t)^2}},
    \label{eq: GaussT}
\end{align}
where $\sigma(t)^2 = \sigma^2 + \frac{t^2}{4 \sigma^2 m^2}$ and $c= \frac{1}{\sqrt{2 \pi\sigma(t)^2}}$.

It can be seen from Eq.~(\ref{eq: GaussT}) that $|\bra{x} e^{-i H t} \ket{\psi}|^2$ is a stationary Gaussian wave packet $e^{-\frac{(x+2b m \sigma^2)^2}{2 \sigma(t)^2}}$ multiplying an exponential factor $e^{2 b m x}$. Due to this exponential factor $e^{2 b m x}$, the peak $x_p$ of the wave packet is not stationary,
\begin{align}
    x_p(t) = 2 b m \left[\sigma(t)^2 - \sigma^2 \right] = \frac{b \, t^2}{2 \sigma^2 m},
    \label{eq: peakX}
\end{align}
satisfying
\begin{align}
    \frac{d}{d x}\vert_{x=x_p} \left( |\bra{x} e^{-i H t} \ket{\psi}|^2 \right) = 0.
    \label{eq: peakEq}
\end{align}
The velocity of the wave packet peak is 
\begin{align}
    v_p(t) = \frac{d x_p}{d t} = \frac{b \, t}{\sigma^2 m}.
    \label{eq: peakV}
\end{align}
Note that if $b=0$ (the Hermitian case), the wave packet peak is stationary and the velocity of the wave packet peak is zero. Otherwise, the wave packet peak is uniformly accelerated. It is also worth noting that the norm square $\| e^{-i H t} \ket{\psi} \|^2$ of the state is $e^{2 b^2 m^2 \left[\sigma(t)^2 - \sigma^2\right]}$, which grows over time. The increasing of the norm square $\| e^{-i H t} \ket{\psi} \|^2$ also comes from from the exponential factor $e^{2 b m x}$, which causes the wave packet peak to grow in height as it approaches the boundary.

Nonzero $b$ and the dependence of $t$ in $\sigma(t)$ lead to a nonzero $v_p(t)$ in Eq.~(\ref{eq: peakV}), and the additional moving velocity deponds on both the magnitude of NHSE and the speed of wave packet spreading. Hence, it can be concluded that the additional moving velocity is the result of NHSE and wave packet spreading.

\subsection{Incident and reflected velocity}
Now, consider the initial Gaussian wave packet with a nonzero central momentum $k_0$
\begin{align}
    \braket{x | \psi} = \frac{1}{(2 \pi \sigma^2)^{\frac{1}{4}}} e^{-\frac{x^2}{4 \sigma^2} + i k_0 x}. 
\end{align}
Under the time evolution governed by the Hermitian Hamiltonian $\bar{H}$, it has a nonzero initial velocity 
\begin{align}
    v_0 = \frac{\partial \bar{H}(k)}{\partial k} \vert_{k_0} = \frac{k_0}{m}.
\end{align}
The time evolution of $\ket{\psi}$ can be calculated using
\begin{align}
    \bra{x} e^{-i H t} \ket{\psi} & = e^{b m x} \bra{x} e^{-i \bar{H} t} \ket{S^{-1} \psi} \notag \\
    & = e^{b m x} \int d k \, \braket{x | k}  e^{-i \bar{H}(k) t} \braket{k | S^{-1} \psi}
    \label{eq: evok0}
\end{align}
and its probability density is
\begin{align}
    |\bra{x} e^{-i H t} \ket{\psi}|^2 \propto e^{-\frac{\left[ x- v_0 t - x_p(t) \right]^2}{2 \sigma(t)^2}},
\end{align}
where $\sigma(t)^2 = \sigma^2 + \frac{t^2}{4 \sigma^2 m^2}$ and $x_p(t) = 2 b m \left[\sigma(t)^2 - \sigma^2 \right]$. This expression is valid before the wave packet reaches the boundary, and $v_0 t$ in the expression is the result of the Hermitian dynamics $e^{-i \bar{H} t} \ket{S^{-1} \psi}$, whereas $x_p(t)$ is the result of the exponential factor $e^{b m x}$ in Eq.~(\ref{eq: evok0}). From above, the velocity before the wave packet reaches the boundary is 
\begin{align}
    v_{\textrm{in}}(t) = v_0 + 2 b m \frac{d \sigma(t)^2}{dt} = v_0 + \frac{b}{m \sigma^2} t.
    \label{eq: vIn}
\end{align}

Next, let us consider the time evolution after the wave packet is bounced back by the boundary, assuming the boundary is located at $x_b$, we assert that
\begin{align}
    |\bra{x} e^{-i \bar{H} t} \ket{S^{-1} \psi}|^2 \propto e^{-\frac{( x - 2 x_b + v_0 t - 2b m \sigma^2 )^2}{2 \sigma(t)^2}},
\end{align}
where we fix the origin for the time to be the moment the wave packet begins to move.
It can be obtained by the following derivation. Before the wave packet reaches the boundary, 
\begin{align}
    e^{-i \bar{H} t} \ket{S^{-1} \psi} = \int dk \ket{k} \bra{k} e^{-i \bar{H} t} \ket{S^{-1} \psi} 
\end{align}
is a linear combination of a series of $\ket{k}$ states. Assume the boundary is at $x_b$, after the full reflection, these $\ket{k}$ states are reflected to $\ket{\tilde{k}}$ states satisfying 
\begin{align}
    \braket{x | \tilde{k}} = e^{i(-k)(x-x_b) + i (k x_b + \pi) }.
\end{align}
During the reflection process, each $\ket{k}$ state acquires a phase shift of $\pi$. Hence, after the full reflection, 
\begin{align}
    e^{-i \bar{H} t} \ket{S^{-1} \psi} = \int dk \ket{\tilde{k}} \bra{k} e^{-i \bar{H} t} \ket{S^{-1} \psi}
\end{align}
and 
\begin{align}
    \bra{x} e^{-i \bar{H} t} \ket{S^{-1} \psi} & = \int dk \braket{x | \tilde{k}} \bra{k} e^{-i \bar{H} t} \ket{S^{-1} \psi} \notag \\
    & = - \int \frac{dk}{2 \pi} e^{i k (2 x_b- x)} \bra{k} e^{-i \bar{H} t} \ket{S^{-1} \psi}.
    \label{eq: afterR}
\end{align}
It can be seen from Eq.~(\ref{eq: afterR}) that compared to the time evolution before the reflection, Eq.~(\ref{eq: afterR}) just substitutes $x$ with $2 x_b -x$ and multiplies a minus sign. Thus, after the full reflection,
\begin{align}
    |\bra{x} e^{-i \bar{H} t} \ket{S^{-1} \psi}|^2 & \propto e^{-\frac{(2 x_b - x + 2b m \sigma^2 - v_0 t)^2}{2 \sigma(t)^2}} \notag \\
    & = e^{-\frac{(x- 2 x_b + v_0 t - 2 b m \sigma^2)^2}{2 \sigma(t)^2}}. \notag
\end{align}
By Eq.~(\ref{eq: evok0}), after the full reflection,
\begin{align}
    |\bra{x} e^{-i H t} \ket{\psi}|^2 \propto e^{2 b m x} e^{-\frac{( x- 2 x_b + v_0 t - 2b m \sigma^2)^2}{2 \sigma(t)^2}}.
\end{align}
The peak of the wave packet $\tilde{x}_p(t)$ satisfying $\frac{d}{dx} \vert_{\tilde{x}_p} |\bra{x} e^{-i H t} \ket{\psi}|^2 =0$ is 
\begin{align}
    \tilde{x}_p(t) = 2 b m \left[ \sigma(t)^2 + \sigma^2 \right] + 2 x_b - v_0 t.
\end{align}
Hence, the velocity of the wave packet after the full reflection is 
\begin{align}
    v_{\textrm{ref}}(t) = -v_0 + 2 b m \frac{d \sigma(t)^2}{dt} = -v_0 + \frac{b}{m \sigma^2} t.
    \label{eq: vOut}
\end{align}
From Eq.~(\ref{eq: vIn}) and Eq.~(\ref{eq: vOut}), it can be seen that $v_{\textrm{in}} \neq -v_{\textrm{ref}}$, which is analogous to the inelastic scattering process in classical mechanics.

\section{Simulation method}
In our simulation, we represent differential operators in terms of finite dimensional matrices using the finite difference method. The Laplace operator can be approximately written as \cite{sauer2012numerical}
\begin{align}
    \nabla^2 \psi_n \approx \frac{\psi_{n+1}+\psi_{n-1}-2\psi_{n}}{(\Delta x)^2}.
\end{align}
Under the Dirichlet boundary condition (taking $\psi_{-1}=\psi_{N+1}=0$), the Laplace operator can be written as
\begin{align}
    \nabla^2 = \frac{1}{(\Delta x)^2} \begin{pmatrix}
        -2 & 1 & 0 & \cdots & \cdots & 0 \\
        1 & -2 & 1 & 0  & \cdots & 0 \\
        0 & 1 & -2 & 1 & \cdots & 0 \\
        \vdots & \ddots & \ddots & \ddots & \ddots & \vdots \\
        0 & \cdots & 0 & 1 & -2 & 1 \\
        0 & \cdots & \cdots & 0 & 1 & -2
    \end{pmatrix}_{N \times N} .
\end{align}
$\nabla$ can be approximately written as (the two-point forward finite difference formula) \cite{sauer2012numerical}
\begin{align}
    \nabla \psi_n \approx \frac{\psi_{n+1}-\psi_{n}}{\Delta x}.
\end{align}
Its matrix form is 
\begin{align}
    \nabla = \frac{1}{\Delta x} \begin{pmatrix}
        -1 & 1 & 0 & \cdots & \cdots & 0 \\
        0 & -1 & 1 & 0  & \cdots & 0 \\
        0 & 0 & -1 & 1 & \cdots & 0 \\
        \vdots & \ddots & \ddots & \ddots & \ddots & \vdots \\
        0 & \cdots & 0 & 0 & -1 & 1 \\
        0 & \cdots & \cdots & 0 & 0 & -1
    \end{pmatrix}_{N \times N} .
\end{align}
The number of the lattice points $N = \frac{L}{\Delta x}$. The accuracy of the finite difference method improves when reducing $\Delta x$ and increasing the number of the mesh cells in the simulation. In our simulation, compared to the length $L =10$, we use $\Delta x= 0.01$, in other words, there are $1000$ lattice points.

\section{Non-wave packet reflection}
\begin{figure} 
    \centering
    \includegraphics[width=0.5\columnwidth]{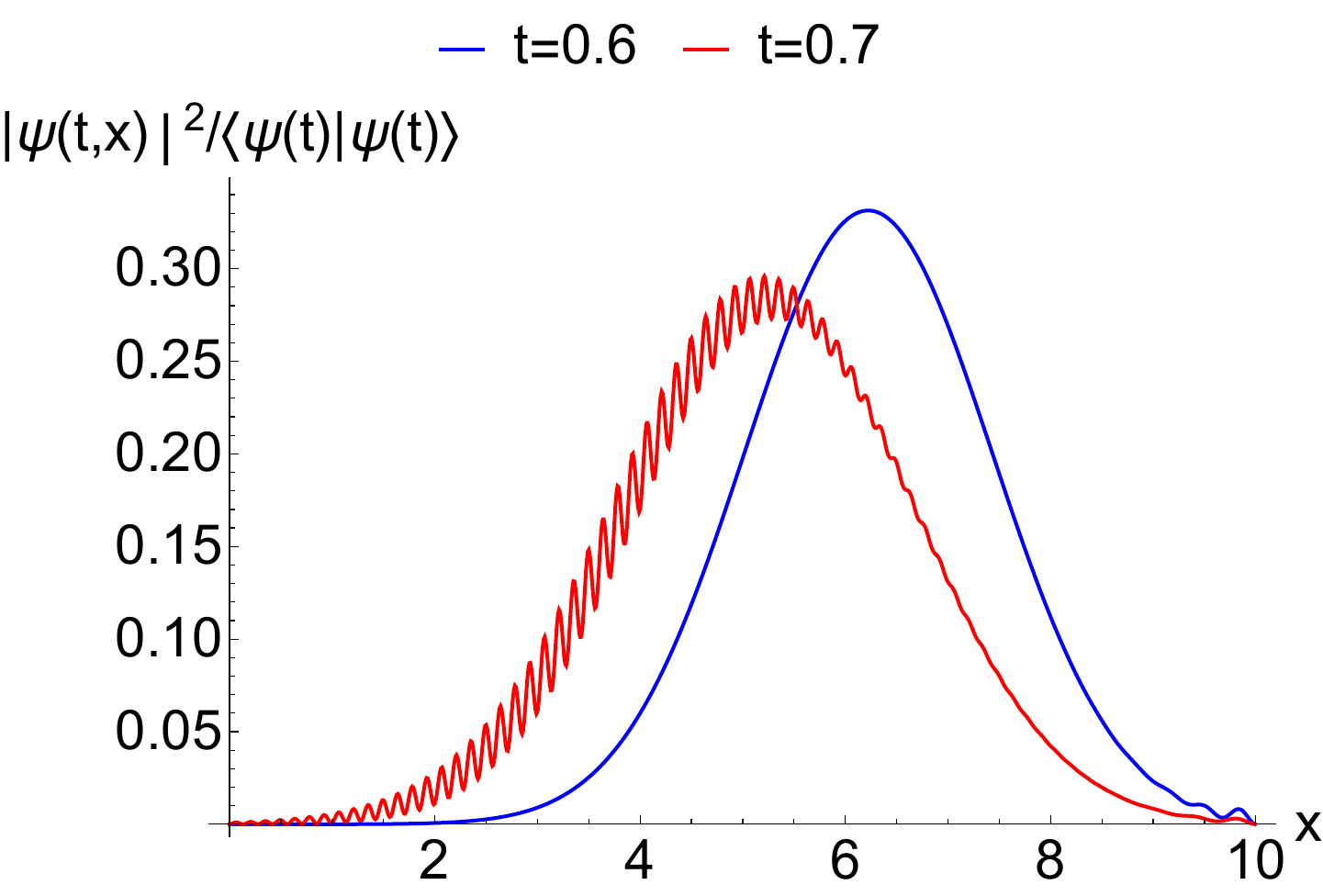}
    \caption[]{At $t = 0.6$ and $0.7$, the probability density (normalized) is depicted in blue and red, respectively.}
    \label{fig: nonWavePacket}
\end{figure}

In Fig.~3b of the main text, the $v(t)$ relation after $t=0.6$ is not taken into count since after $t=0.6$, the wave packet width is large enough in comparison to the length $L$ of the system, and extra non-wave packet reflection occurs, as shown in Fig.~\ref{fig: nonWavePacket}. This process is due to the interference of the incident wave and reflected wave, and is not of our interest.

\section{Snapshots of the wave meeting process}
\begin{figure} 
    \subfigure[]{
        \centering
        \includegraphics[width=0.32\columnwidth]{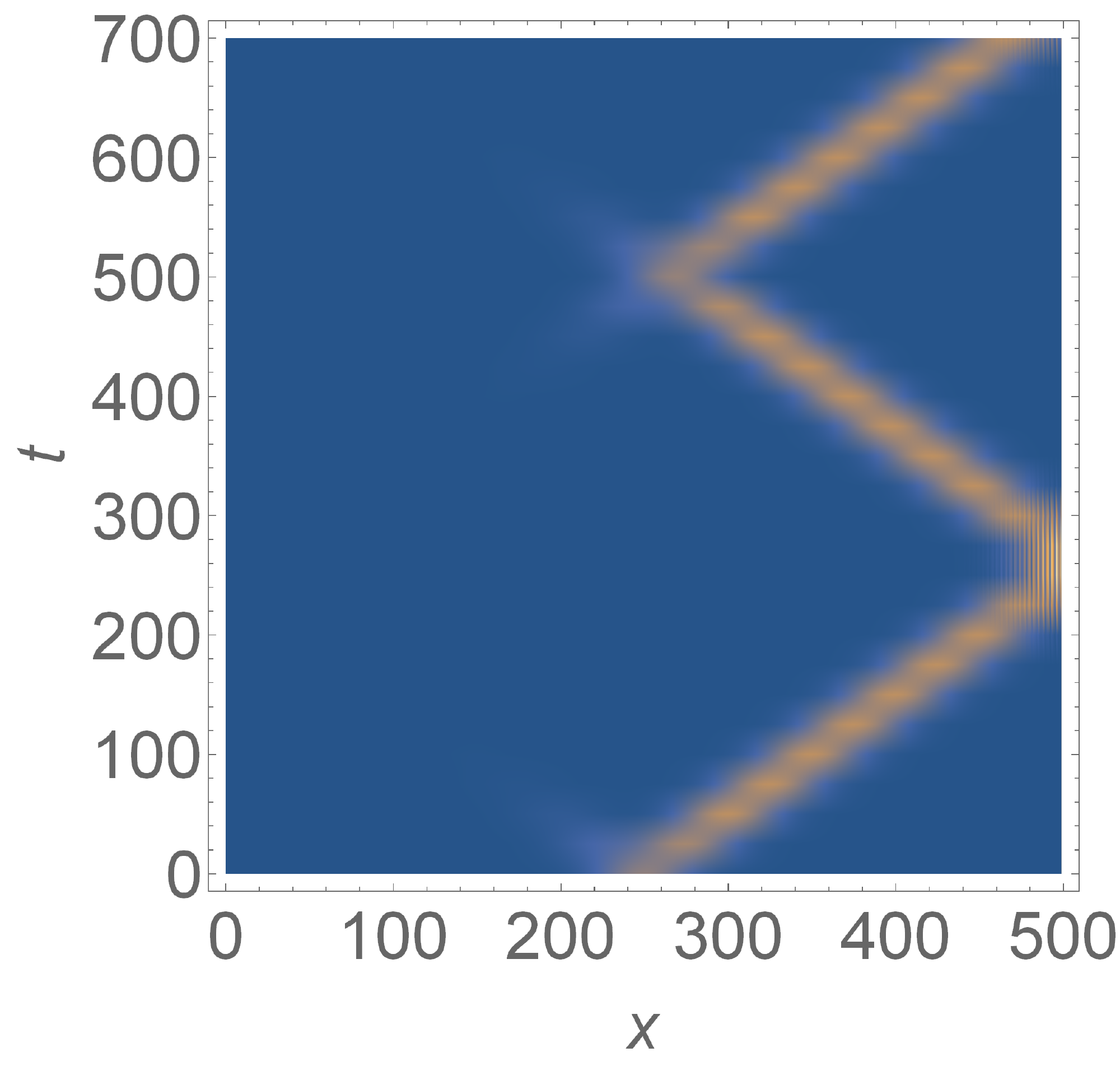}
    }
    \subfigure[]{
        \centering
        \includegraphics[width=0.45\columnwidth]{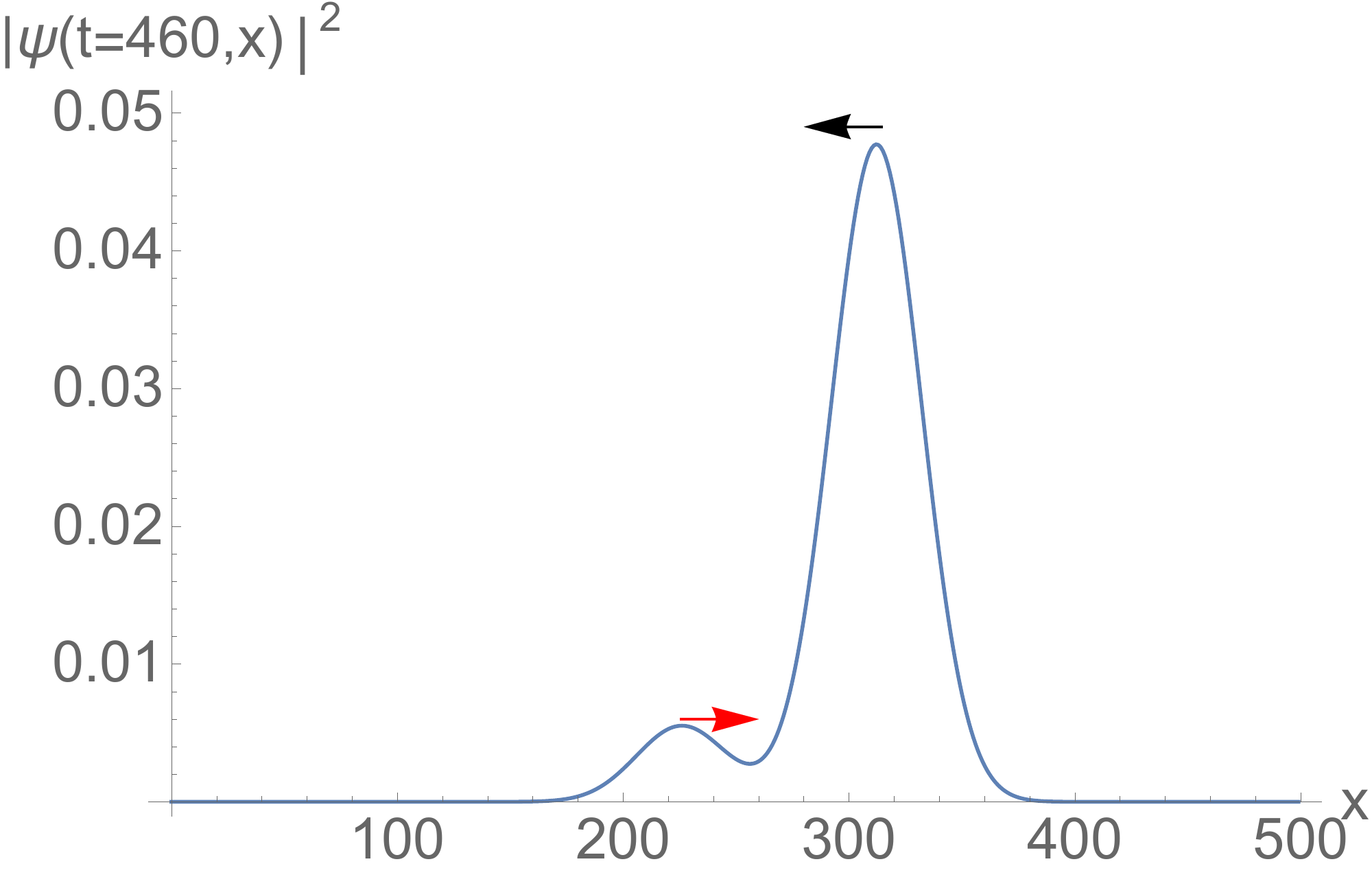}
    }
    \subfigure[]{
        \centering
        \includegraphics[width=0.45\columnwidth]{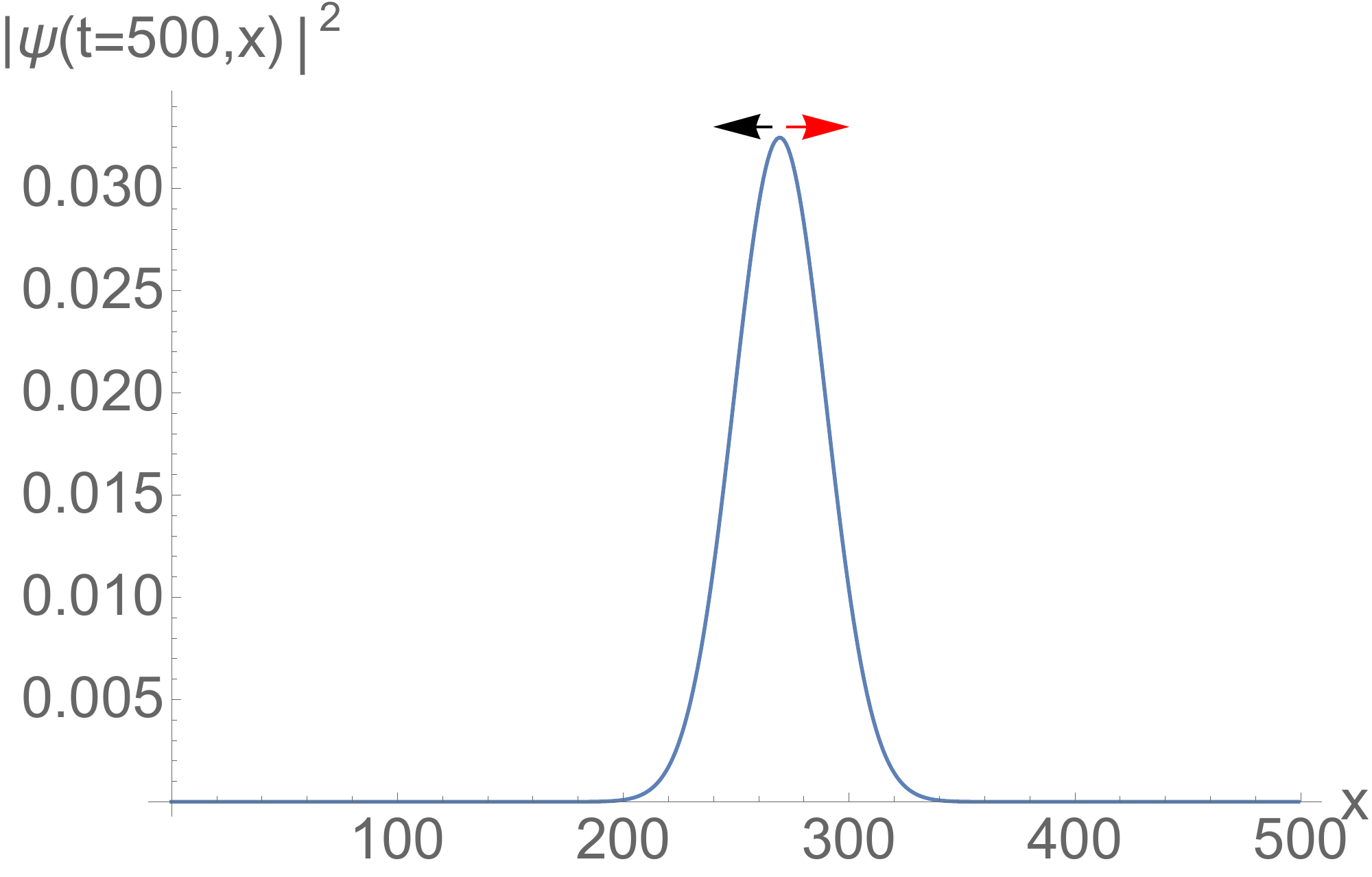}
    }
    \subfigure[]{
        \centering
        \includegraphics[width=0.45\columnwidth]{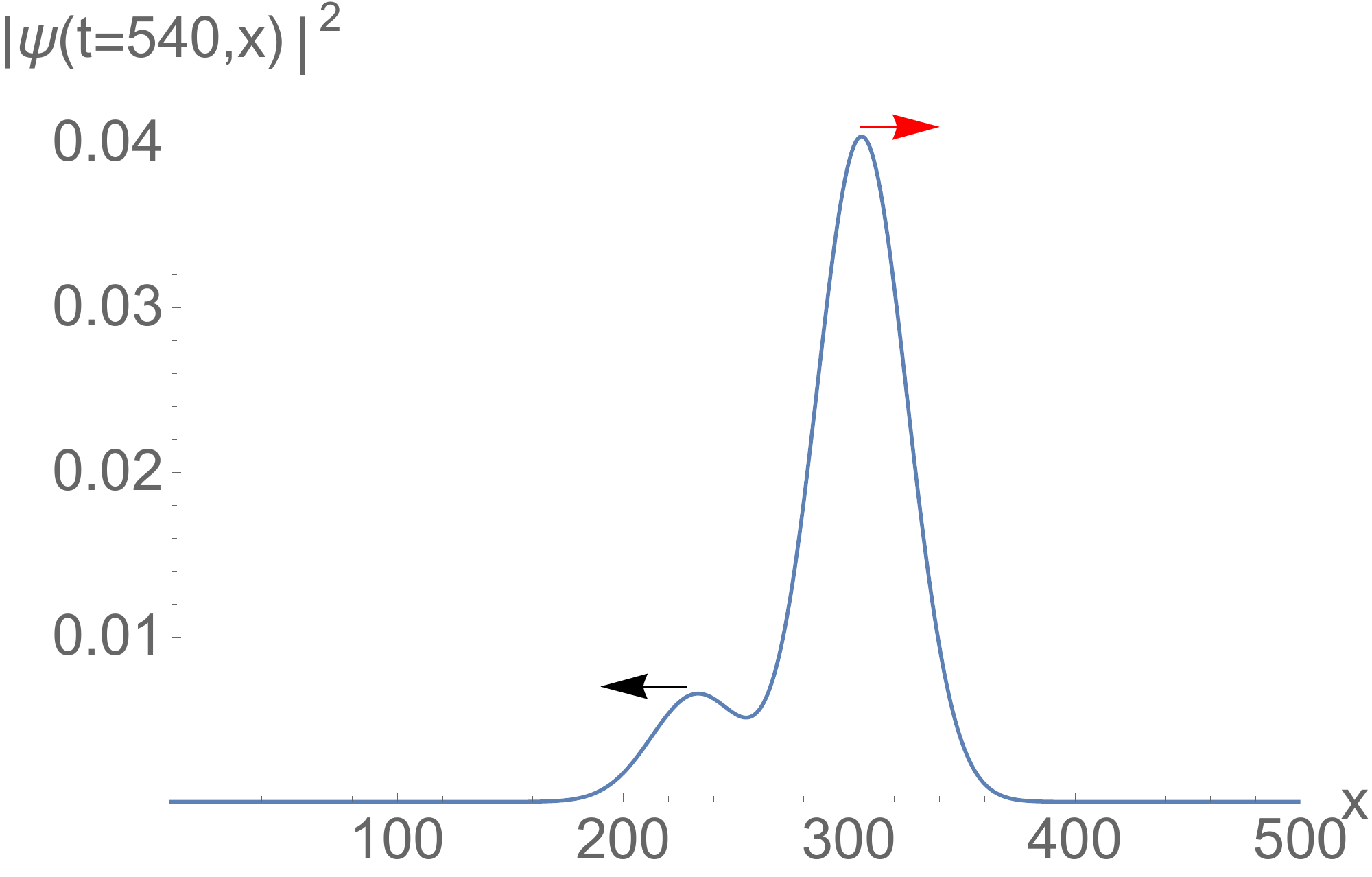}
    }
    \caption[]{Wave meeting process in the non-Hermitian SSH model. In the simulation, $k_0=2$ and $\gamma=-0.05$. (a) Numerical simulation of the probability distribution $|\braket{x | \psi(t)}|^2$ (normalized) as a function of $x$ and $t$. It can be seen that two wave packets meet each other at $t=500$. (b) At $t=460$, the probability distribution before two wave packets meet each other. (c) The probability distribution when two wave packets meet each other at $t=500$. (d) At $t=540$, the probability distribution after two wave packets meet. In (b), (c), and (d), the directions of the wave packet velocities are represented as arrows.}
    \label{fig: wave_meet}
\end{figure}

In Fig.~\ref{fig: wave_meet}, we show snapshots of a wave meeting process in the non-Hermitian SSH model.

\section{The effect of wave packet spreading on the reflection}
Eq.~(16) of the main text shows that the velocity of the additional motion of the wave packet depends on both the magnitude of NHSE and the speed of the wave packet spreading. By Eq.~(17) of the main text, as a direct consequence, the ratio of the reflected velocity to the incident velocity of the wave packet depends both the magnitude of NHSE and the speed of the wave packet spreading. In the study of the dynamic skin effect, the extreme situation that the reflected wave packet remaining at the boundary and the conditions that lead to this is of our particular interests. It is discovered that when the wave packet encounters the boundary, whether it stays at the boundary depends on both the magnitude of NHSE and the speed of the wave packet spreading. We now demonstrate how the wave packet reflection in non-Hermitian SSH models is impacted by wave packet spreading. 

\begin{figure} 
    \centering 
    \subfigure[]{
        \centering
        \includegraphics[width=0.45\columnwidth]{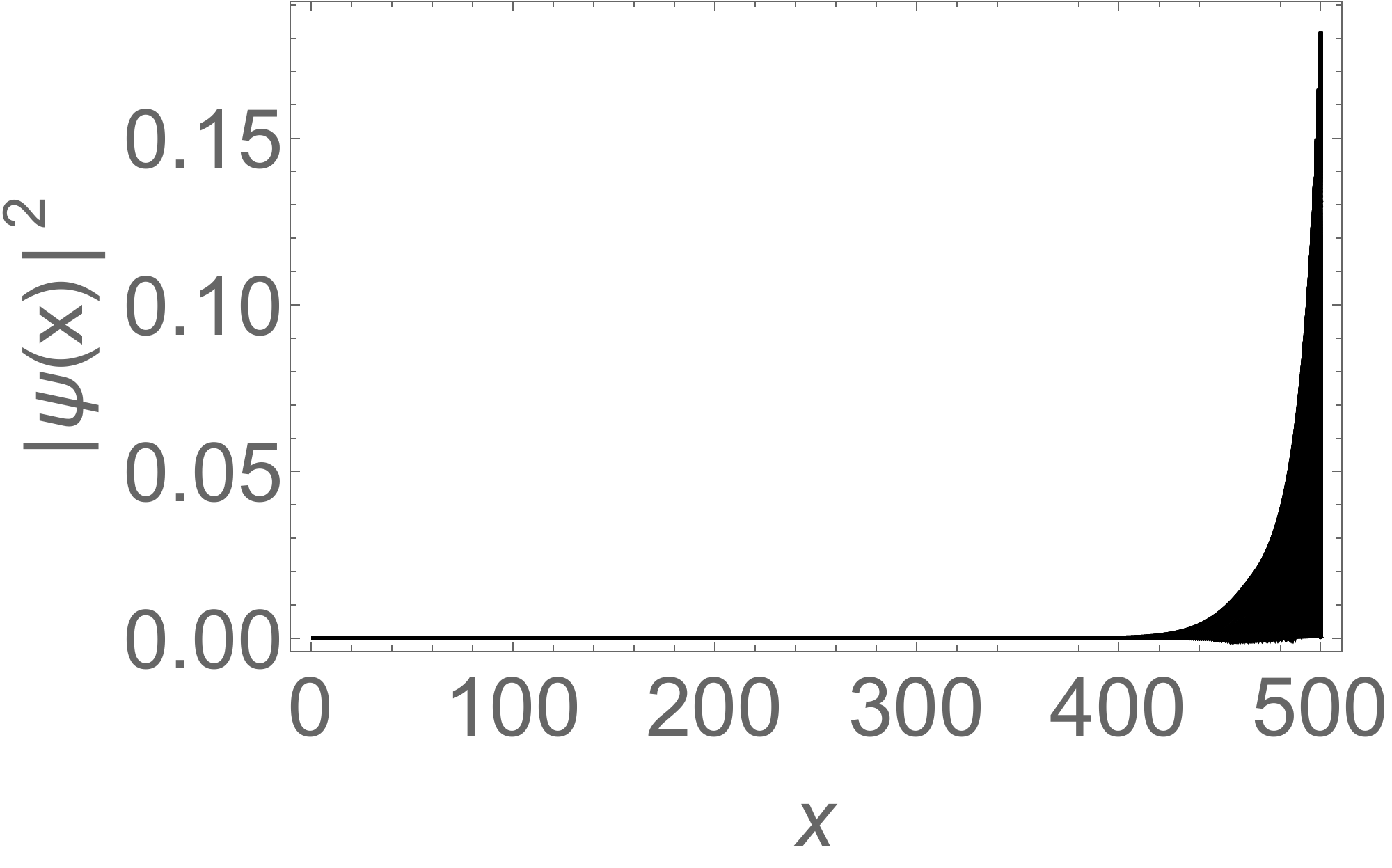}
    }
    \subfigure[]{
        \centering
        \includegraphics[width=0.45\columnwidth]{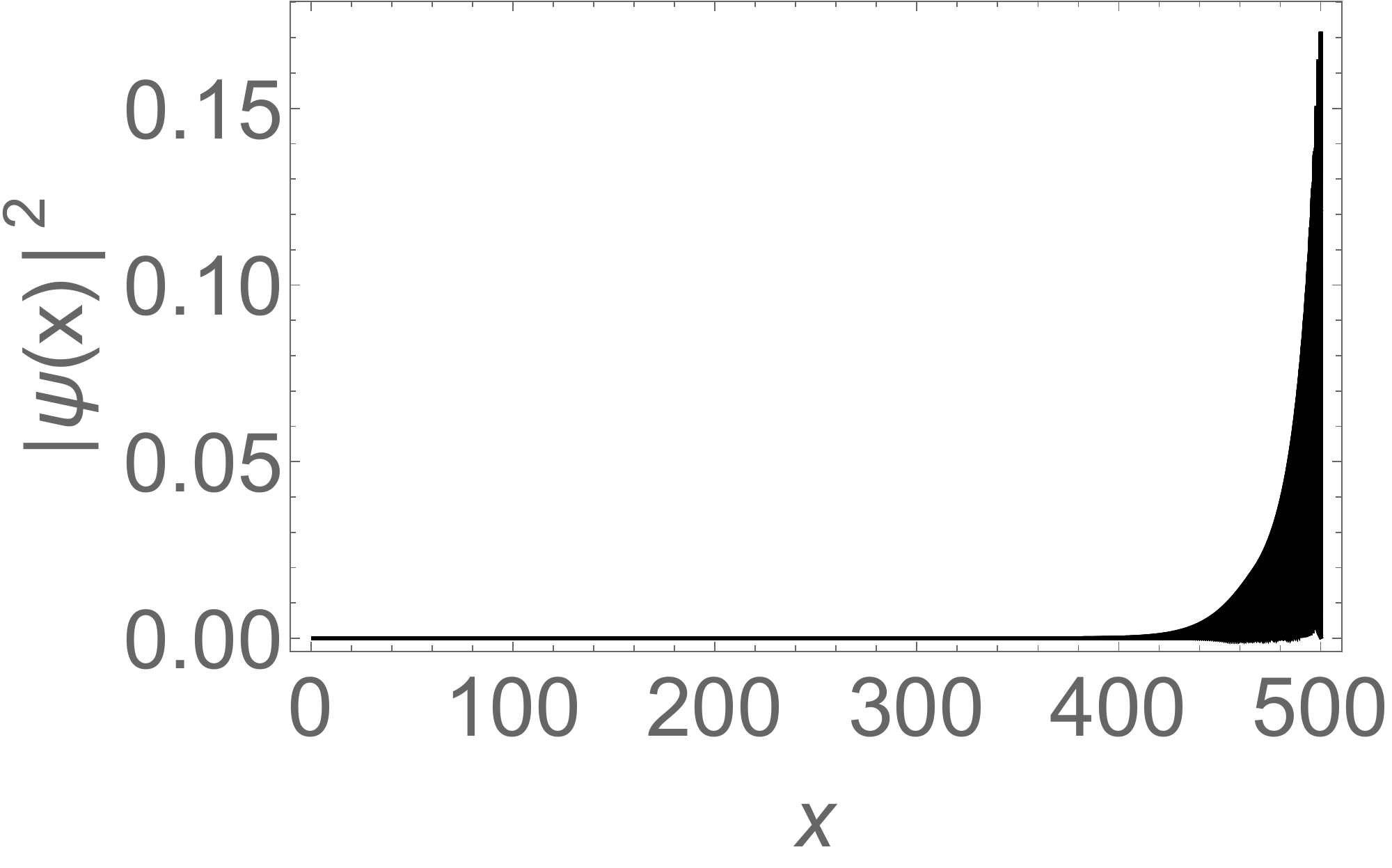}
    }
    \subfigure[]{
        \centering
        \includegraphics[width=0.45\columnwidth]{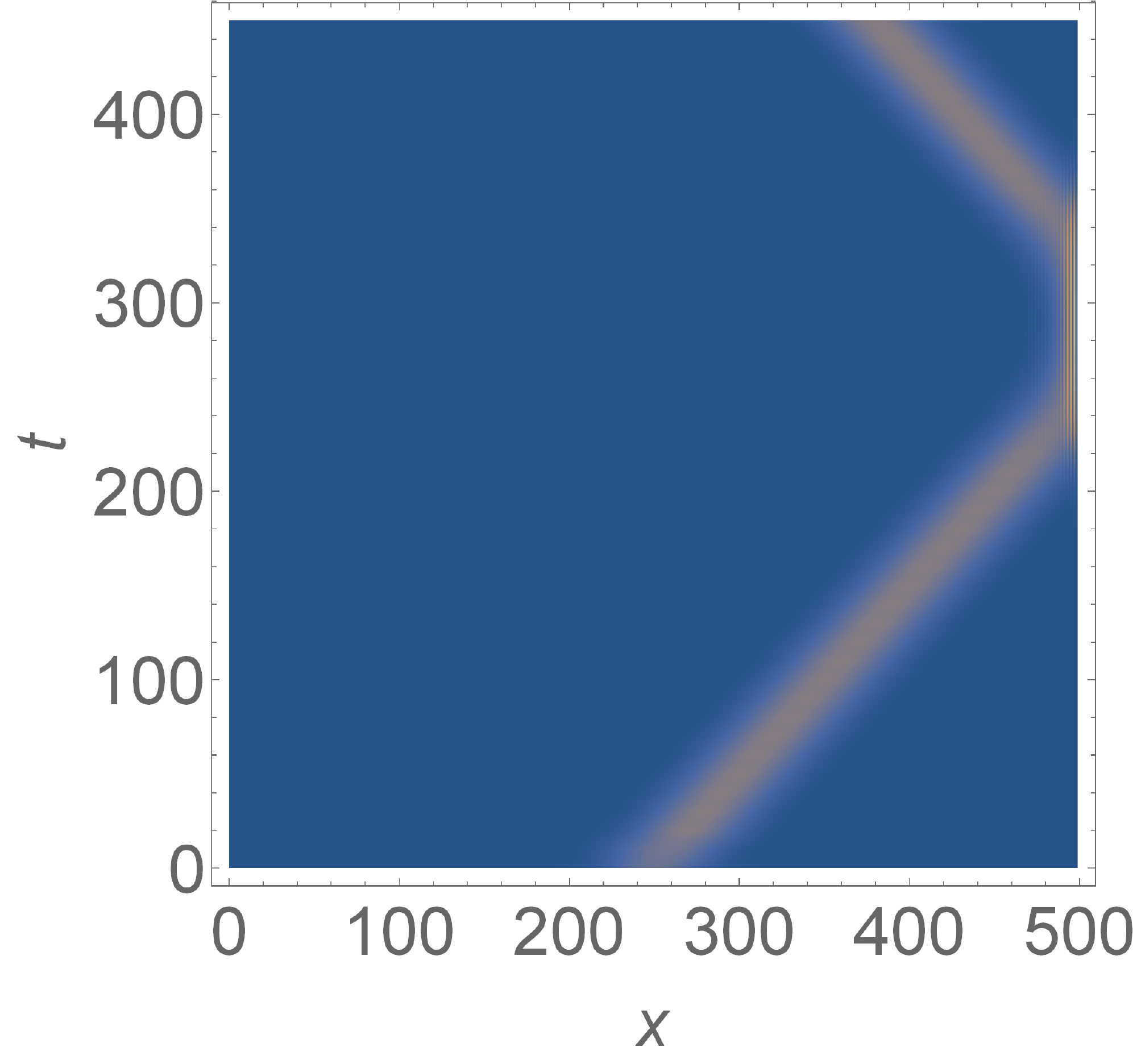}
    }
    \subfigure[]{
        \centering
        \includegraphics[width=0.45\columnwidth]{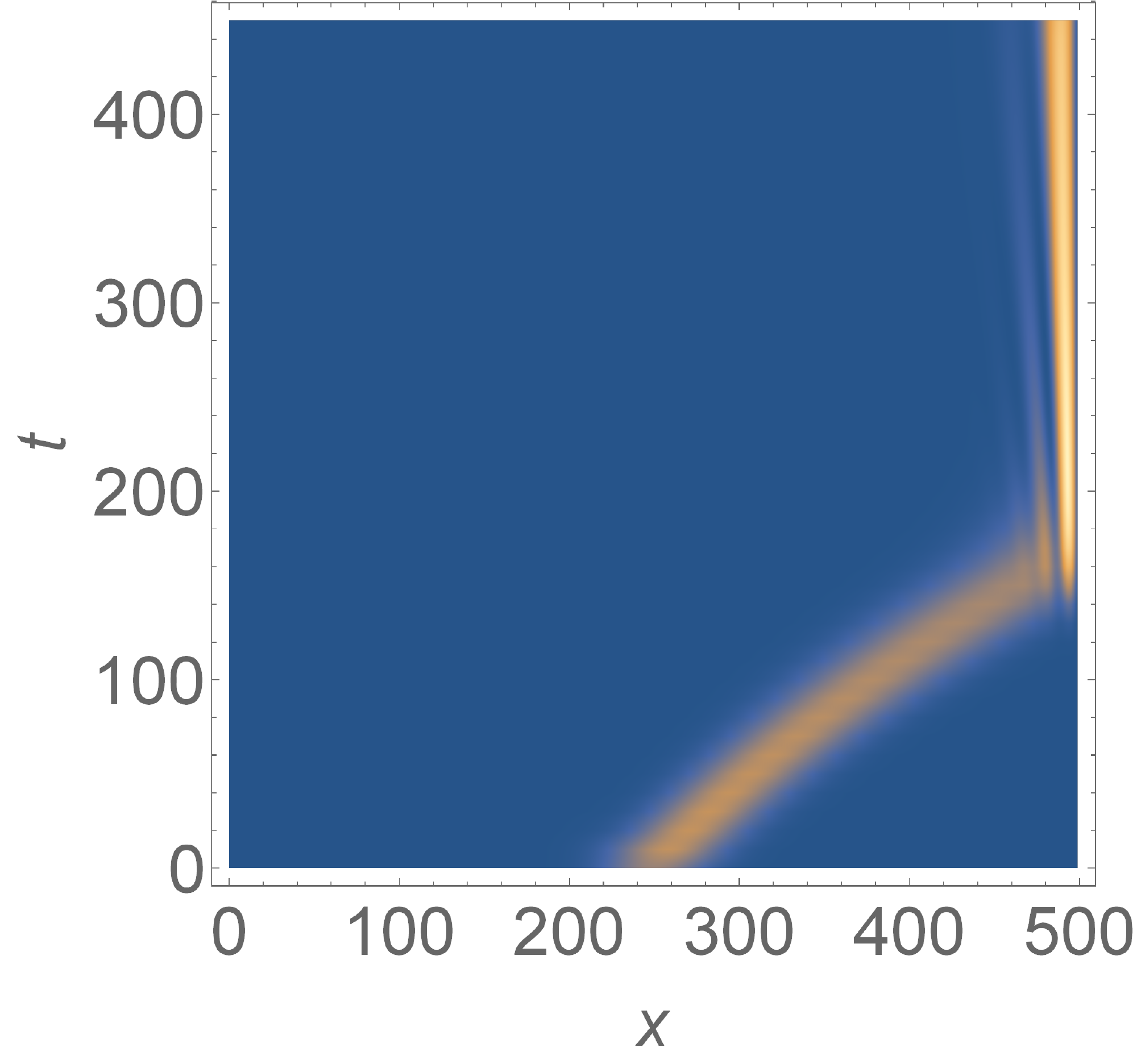}
    }
    \caption[]{The collections of non-Hermitian skin modes and the time evolution of the wave packets in non-Hermitian SSH models with different $t_2$. In these two models, $t_1=20$, $\gamma=-2$ and $L=500$, the initial velocity $v_{0,-}$ of the wave packet is $1$, and the standard derivation $\sigma$ of the initial wave packet is $20$. (a),(c) represent the model with $t_2 = 1$. (b),(d) represent the model with $t_2=10$. (a),(b) shows the collections of non-Hermitian skin modes. (c),(d) shows the the time evolution of the wave packets of corresponding models.}
    \label{fig: sameSkin}
\end{figure}

In the non-Hermitian SSH model, the magnitude of NHSE (localization length of skin modes) is completely controlled by parameters $t_1$ and $\gamma$. Hence, if one only varies $t_2$, the speed of wave packet spreading is changed without changing the localization length of skin modes. Consider the wave packet with the energy dispersion $E_{-}(k)$, by Eq.~(17) in the main text, $v_{\text{ref}}(t) \geqslant 0$ requires $v_{0,-} = \frac{\partial E_{-}(k)}{\partial k} \vert_{k_0} \leqslant 2 \ln\left( \sqrt{|\frac{t_1 - \gamma/2}{t_1 + \gamma/2 }|} \right) \frac{d \, \sigma(t)^2}{d t}$. Therefore, when the wave packet encounters the boundary, the wave packet spreading slowly (with small $t_2$) will be reflected by the boundary while the wave packet spreading quickly (with large $t_2$) will stay at the boundary. In Fig.~\ref{fig: sameSkin}, we show the time evolution of the wave packets with the fixed initial velocity $v_{0,-}$ in the non-Hermitian SSH model with varying $t_2$. It can be seen that the wave packets can either be reflected by the boundary or stay at the boundary although the localization length of skin modes is the same in these two models.

\section{Turn on non-Hermiticity near the boundary}
The incident velocity and reflected velocity of the wave packet are (Eq.~(17) in the main text)
\begin{align}
    v_{\textrm{in}}(t) & = \frac{\partial E(k)}{\partial k}\vert_{k_0}  + v_p(t) , \notag \\
    v_{\textrm{ref}}(t) & = \frac{\partial E(k)}{\partial k}\vert_{k_1} + v_p(t) .
    \label{eq: in_ref}
\end{align}
The same sign of the additional velocity $v_p(t)$ caused by non-Hermiticity is the origin of the dynamic skin effect. The additional velocity $v_p(t)$ can be turned on by local gain and loss, hence, the dynamic skin effect still exists when one only turns on non-Hermiticity close to the boundary. 

\begin{figure}
    \centering
    \includegraphics[width=0.45\columnwidth]{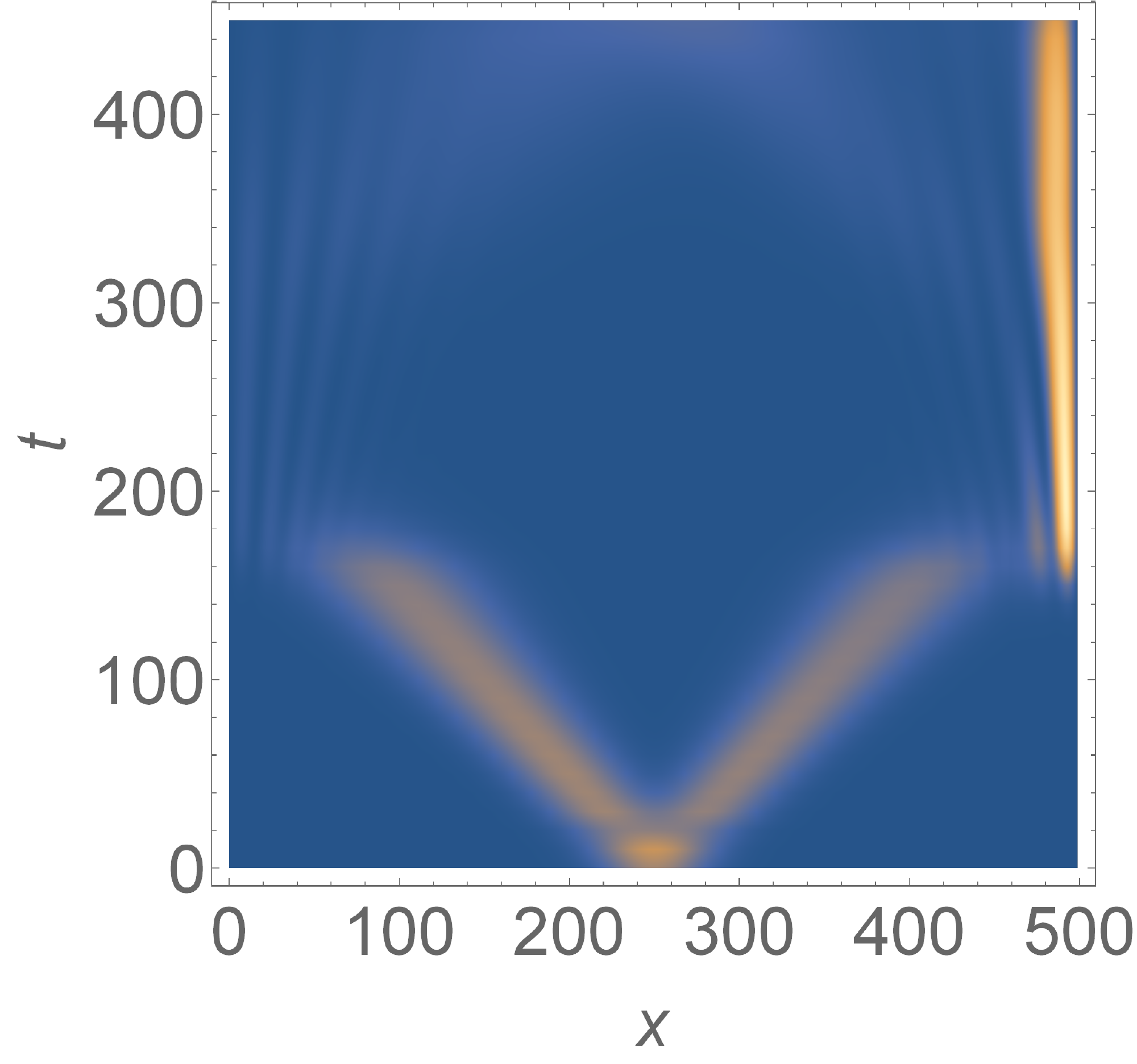}
    \caption[]{The extreme case of the dynamic skin effect that the reflected wave packet stays at the boundary and only on-site gain and loss near the right end of the one-dimensional chain are turned on. $t_1=20$, $t_2=10$ and $L=500$. $\gamma=-2$ in the 40 unit cells on the far right while $\gamma=0$ in the bulk.}
    \label{fig: local_nonHerm}
\end{figure}

Consider the following non-Hermitian Bloch Hamiltonian realized by on-site gain and loss, $H(k) = (t_1 + t_2 \cos k) \sigma_x + (t_2 \sin k + i \frac{\gamma}{2}) \sigma_z$ with $i \frac{\gamma}{2} \sigma_z$ corresponding to gain(loss) at A(B) sublattice. We only turn on on-site gain and loss with nonzero $\gamma$ near the right end of the one-dimensional chain and $\gamma$ is zero in the bulk. By Eq.~(\ref{eq: in_ref}), the dynamic skin effect still exists in this system which is shown in Fig.~\ref{fig: local_nonHerm}.

\bibliography{main}